\pdfoutput=1
\documentclass[11pt,a4paper]{article}
\usepackage{jheppub}
\usepackage[caption=false]{subfig}
\usepackage{multirow}
\usepackage{doi}

\newcommand\Lag{\mathcal{L}}
\newcommand\Ham{\mathcal{H}}
\newcommand\p{\partial}

\def\=:{=\hspace{-.7em}\raisebox{1.1ex}{.}\hspace{.1em}\raisebox{-0.2ex}{.}}

\newcommand{\beq}{\begin{eqnarray}}
\newcommand{\eeq}{\end{eqnarray}}
\newcommand{\non}{\nonumber\\}

\newcommand{\bn}{\mathbf{n}}
\newcommand{\bm}{\mathbf{m}}

\newcommand{\ol}[1]{\mkern2.75mu\overline{\mkern-2.75mu#1\mkern-1.5mu}\mkern1.5mu}
\DeclareMathOperator{\Og}{O}
\DeclareMathOperator{\SO}{SO}
\DeclareMathOperator{\SU}{SU}
\newcommand{\NLSM}{NL$\sigma$M}
\newcommand{\CP}{\mathbb{C}P}
\renewcommand{\i}{\mathrm{i}}

\newcommand\zb{\bar{z}}
\newcommand\Hb{\ol{H}}
\newcommand\Bb{\ol{B}}
\newcommand\Eb{\ol{E}}
\newcommand\Gb{\ol{G}}
\newcommand\Ib{\ol{I}}
\newcommand\Jb{\ol{J}}
\newcommand\Qb{\ol{Q}}
\newcommand\Rb{\ol{R}}
\newcommand\Sb{\ol{S}}
\newcommand\Tb{\ol{T}}

\newcommand\psib{\bar{\psi}}
\newcommand\pib{\bar{\pi}}
\newcommand\wb{\bar{w}}
\newcommand\cc{\textrm{c.c.}}
\newcommand\Ktilde{\widetilde{K}}
\newcommand\Htilde{\widetilde{H}}
\newcommand\Hbtilde{\widetilde{\Hb}}

\newtheorem{theorem}{Theorem}
\newtheorem{corollary}[theorem]{Corollary}

\newtheorem{lemma}[theorem]{Lemma}
\newtheorem{remark}[theorem]{Remark}
\newtheorem{proposition}[theorem]{Proposition}

\begin{document}

\title{Reducing the $\Og(3)$ model as an effective field theory}

\author{Sven Bjarke Gudnason$^1$,}
\affiliation{$^1$Institute of Contemporary Mathematics, School of
  Mathematics and Statistics, Henan University, Kaifeng, Henan 475004,
  P.~R.~China}
\emailAdd{gudnason(at)henu.edu.cn}
\author{Muneto Nitta$^2$}
\affiliation{$^2$Department of Physics, and Research and Education
  Center for Natural Sciences, Keio University, Hiyoshi 4-1-1,
  Yokohama, Kanagawa 223-8521, Japan}
\emailAdd{nitta(at)phys-h.keio.ac.jp}

\abstract{
We consider the $\Og(3)$ or $\mathbb{C}P^1$ nonlinear sigma model
as an effective field theory in a derivative expansion, with
the most general Lagrangian that obeys $\Og(3)$, parity and Lorentz
symmetry.
We work out the complete list of possible operators (terms) in the
Lagrangian and eliminate as many as possible using integrations by
parts.
We further show at the four-derivative level, that the theory can be
shown to avoid the Ostrogradsky instability, because the dependence on
the d'Alembertian operator or so-called box, can be eliminated by a
field redefinition.
Going to the six-derivative order in the derivative expansion, we show
that this can no longer be done, unless we are willing to sacrifice
Lorentz invariance. By doing so, we can eliminate all dependence on
double time derivatives and hence the Ostrogradsky instability or
ghost, however, we unveil a remaining dynamical instability that takes
the form either as a spiral instability or a runaway instability and
estimate the critical field norm, at which the instability sets off. 
}


\maketitle

\section{Introduction and summary}

Effective field theories (EFTs) at low energies are very useful tools
in classical and quantum field theories for formulating the most 
important impact of the full theory in terms of the low-energy
variables (fields) valid at low energies up to the scale of the
lightest field that has been integrated out, see
e.g.~ref.~\cite{Manohar:2018aog} for a review.
Nowadays, it is even a commonly used approach for parametrizing our
ignorance of new physics beyond the standard model and in that context
it is called the standard model effective field theory (SMEFT)
\cite{Brivio:2017vri}.

A more traditional low-energy EFT is the chiral Lagrangian as the
low-energy theory of the strong interactions, which is a step more
radical than the SMEFT, since the original fields, viz.~quark and
gluon fields, of quantum chromodynamics (QCD) are absent in the chiral
Lagrangian theory and the lightest fields are the Nambu-Goldstone
bosons of chiral symmetry breaking, namely the pions
\cite{Scherer:2002tk,Epelbaum:2008ga,Machleidt:2011zz}.
In chiral perturbation theory, the pions and other vector mesons are
coupled to a nucleon field and the expansion is made order-by-order in
the so-called chiral order which counts the number of derivatives or
powers of quark masses, where a quark mass counts as two derivatives.

The leading-order term in the chiral Lagrangian, in the mesonic
sector, is the kinetic term of the pions, which is written as a
kinetic term of an $\SU(2)$-valued field containing three pions and an
auxiliary field, traditionally called $\sigma$. The latter auxiliary
field allows for a nonlinear constraint on the $\SU(2)$-valued field
of the chiral Lagrangian, such that its determinant is manifestly
equal to unity.
This constraint in turn induces a nonvanishing curvature for the
metric on the target space of the pions.
For historic reasons and due to the choice of $\sigma$ as the symbol
for the auxiliary field, such kind of theory, has henceforth been
coined a nonlinear sigma model (\NLSM{}).

Due to a low-energy theorem established by Weinberg
\cite{Weinberg:1978kz}, the effective field theory can be considered
as a meaningful theory, despite the fact that it is nonrenormalizable
and the systematic scheme usually utilized works by fixing a chiral
order, $D$, to which calculations are carried out, and only
considering $L\leq\frac{D}{2}-1$ loops in the theory.

A problem, however, naturally occurs at higher orders in a derivative
expansion, namely that there may be more than one derivative acting on
the same field. The simplest example with Lorentz invariance, is the
square of the d'Alembertian operator on a field, $(\square\phi)^2$,
which due to the theorem of Ostrogradsky \cite{Ostrogradsky:1850fid},
see also ref.~\cite{Woodard:2015zca}, must have a linear dependence on at least one
conjugate momentum, since the Hamiltonian corresponding to this Lagrangian contains
$(\p_t^2\phi)^2$. This readily implies that the energy is unbounded
from below and from above and in turn that the theory is plagued by at
least one ghost field.
The necessary assumption for the Ostrogradsky theorem to hold, is that
the Lagrangian is nondegenerate in the double time derivative,
i.e.~$\frac{\p^2\Lag}{(\p\p_t^2\phi)^2}\neq0$.

It is, however, well known that a fundamental theory can be absolutely
sane, whereas upon writing down its effective low-energy theory, for
example by integrating out a massive field, the resulting EFT turns
out to be either nonlocal or in general plagued by the Ostrogradsky
ghost, see e.g.~ref.~\cite[sec.~II.B]{Solomon:2017nlh}.
It can further be argued that the Ostrogradsky ghost is naturally
suppressed by the EFT energy scale, which is the energy scale of the
lightest massive particle that has been integrated out, and hence can
only be dynamically excited at energies of order of the EFT scale,
which is by definition the energy scale where the EFT breaks down
\cite{Solomon:2017nlh}.
In case the theory at hand is in a certain class of asymptotically
free theories, it has further been argued to have an effective mass
that runs to infinity in the ultraviolet limit \cite{Asorey:2020omv}.
It is thus clear that the Ostrogradsky ghost, in a physically sane
theory, should be considered as an artifact of the EFT and not a
physically viable excitation (solution).
Since the ghost by its nature comes with a kinetic term of the wrong
sign, it furthermore implies the loss of unitarity
\cite{Woodard:2015zca}, which is unfortunate for a quantum theory.

A well-known condition on a theory with higher derivatives, is to fine
tune the coefficients of the higher-order derivative terms in such a
way that the Euler-Lagrange equation of motion is of second order.
This is indeed the way the Ostrogradsky ghost or instability is
avoided in Galileon \cite{Nicolis:2008in} and
Horndeski \cite{Horndeski:1974wa} theories of gravity, the Skyrme
model \cite{Skyrme:1961vq} (see ref.~\cite{Gudnason:2017opo} for a higher order generalization), the Faddeev-Skyrme
model \cite{Faddeev:1998eq}, and the baby-Skyrme
model \cite{Piette:1994jt,Piette:1994ug}.
This is of course like cherry-picking the theory in theory space and
cannot be considered as the most general case.

One could contemplate ways to cure the theory in general, for instance
by inventing certain projections that eliminate the unwanted features
of the higher-derivative theory. One such approach has been carried
out in Lee-Wick theory yielding a consistent unitary and
renormalizable higher-derivative theory, however, at the cost of loss
of causality at short distances
\cite{Anselmi:2017lia}. 
It is claimed that the loss of microcausality does not propagate to
larger length scales \cite{Anselmi:2018tmf,Anselmi:2019nie}.

An Ostrogradsky ghost also exists in supersymmetric field theories 
with higher-dimen\-sional operators in general  
\cite{Antoniadis:2007xc,Dudas:2015vka}.
One may eliminate the Ostrogradsky ghost by introducing an auxiliary gauge field to gauge it away, 
which was carried out explicitly in a supersymmetric chiral model
\cite{Fujimori:2016udq} 
and in supergravity \cite{Fujimori:2017rcc}.

It has also been argued that the classical instability of the
Ostrogradsky Hamiltonian does not pose as severe a problem in quantum
theory as in classical physics \cite{Donoghue:2021eto}, although the
analysis performed here seems to require that the theory can be
factorized, in terms of the partition function, into sane theories
with some factors of the partition function being time reversed.
In this case, it can be argued that the quantum field theory is not
suffering directly from the Ostrogradsky instability, although
microcausality is again certainly lost.

Yet another proposal is to restrict the theory to possess a certain
reflection positivity, which essentially is a condition on the kinetic
matrix of an equivalent theory made using more auxiliary fields, all
with a positive sign of their respective kinetic
term \cite{Arici:2017whq}.

Nevertheless, it would be pleasant if it would be possible, already at
the level of the Lagrangian, to eliminate the spurious or non-physical
part of the theory, for instance, by means of finding an appropriate
field redefinition (see e.g.~ref.\cite[sec. 6]{Manohar:2018aog}) that
can absorb the spurious part of the theory \cite{Solomon:2017nlh}. 
This is indeed possible \cite{Solomon:2017nlh} in simple theories,
like the Ostrogradsky extension of the harmonic oscillator
\cite{Woodard:2015zca} and in Galileon modifications of gravity
\cite{Nicolis:2008in} up to mass dimension
11 \cite{Solomon:2017nlh}.
Field redefinitions also play an important role in the renormalization
of effective gauge field theories, see
e.g.~\cite{Gomis:1995jp,Anselmi:2013sx,Quadri:2021syf}.

An important class of effective field theories is based on the \NLSM{}
with the addition of higher derivatives, the Skyrme
model \cite{Skyrme:1961vq}, the Faddeev-Skyrme
model \cite{Faddeev:1998eq}, and the baby-Skyrme
model \cite{Piette:1994jt,Piette:1994ug}, are examples. 
For solitons, like the Skyrmion or the Faddeev-Skyrme knot, the
higher-derivative term is a loophole in Derrick's
theorem \cite{Derrick:1964ww}, preventing the soliton from collapsing.
The latter two models are based on \NLSM{}s with target space $S^2$ or
$\mathbb{C}P^1$, whereas the first has target $S^3\sim\SU(2)$. 

In this paper, we consider the $\Og(3)$ \NLSM{} as an effective theory
and write down the most general higher-order derivative operators and
the goal is to simplify them as much as possible by
integration-by-parts and by using field redefinitions\footnote{Similar 
  considerations have already been made in the context of chiral
  perturbation theory to fourth \cite{Gasser:1983yg}, sixth
  \cite{Bijnens:1999sh}, eighth \cite{Bijnens:2018lez} and even higher
  order \cite{Graf:2020yxt} in derivatives. The analysis there is
  somewhat different, due to the phenomenological theory (chiral
  perturbation   theory) being based on $\SU(2)\times\SU(2)$ symmetry
  and the fact that external gauge and scalar fields are included in
  the expansion, complicating the analysis.}.
We will assume intact $\Og(3)$ symmetry, parity symmetry and Lorentz
(Poincar\'e) symmetry, and only relax the latter symmetry requirement
at the end.
The $\Og(3)$ \NLSM{} is equivalent to the $\mathbb{C}P^1$ \NLSM, by
changing coordinates from a real vector field
$\bn:\mathbb{R}^{d+1}\to\mathbb{R}^3$ to a complex field
$z:\mathbb{R}^{d+1}\to\mathbb{C}$, which is the ordinary Riemann
sphere coordinate.
We will consider the model order-by-order in a derivative expansion,
where each order is suppressed by further powers of a single intrinsic
mass scale of the \NLSM, taken as the scale up to which the EFT is
trustable. 

Our main motivation is two-fold. We would like to see whether it is
possible to get rid of the Ostrogradsky instability in the
$\Og(3)$ \NLSM{} by simplifying and using field redefinitions, since
the Ostrogradsky ghost and corresponding instability is non-physical
and should be considered an artifact of the low-energy effective field
theory.
This is known not to be possible in other theories, like the chiral
Lagrangian, but the only glimmer of hope here could be due
integrability of the $\Og(3)$ \NLSM{}.
The other motivation is to find a minimal formulation of the model,
with an explicit and natural basis of operators describing the model.
It is important in this regard that we clearly define what symmetries,
continuous and discrete, are imposed on the theory for the result,
since changing this assumption will allow further (or fewer) operators
in the theory.

First we contemplate how to construct the most general Lagrangian of
the $\Og(3)$ \NLSM{} with unbroken $\Og(3)$, parity and Lorentz
symmetry.
We establish in Lemma \ref{lemma:even} that with the given symmetry
assumptions, the number of derivatives must be even and hence so does
the suppressing powers of the EFT scale, $\Lambda$. 
At the leading order in the derivative expansion, we establish the
unique Lagrangian, up to an overall constant, in
Theorem \ref{thm:dim2}, Lemma \ref{lemma:topo_term} and
Corollary \ref{coro:dim2unique}.

At the next-to-leading order in the derivative expansion, which we
denote $\Lambda^{-2}$, we exhaust the possibilities of terms or
operators compatible with the symmetry requirements and show in
Theorem \ref{thm:NLSM_Lambda2}, that the most general
$\Og(3)$ \NLSM{} to this order is box-free or d'Alembertian-free,
viz.~it contains no more than a single derivative operator acting on a
field.
This means that the most general theory -- to this order -- is
described by a second-order Euler-Lagrange equation of motion and is
free from the Ostrogradsky instability or ghost.
We utilize integrations by parts and field redefinitions to establish
this result.

At the next-to-next-to-leading order in the derivative expansion,
which we denote $\Lambda^{-4}$, we first write down the complete list
of terms or operators to this order, namely containing six derivative
operators and any number of fields (which can easily be shown to be
6, 4 or 2).
We then establish in Lemma \ref{lemma:Lambda4_intbyparts}, that the
list of operators can be reduced by integrations by parts to a
representative of 10 operators.
We take into account the repercussions of the field redefinition used
at order $\Lambda^{-2}$ to eliminate boxes or d'Alembertian operators,
which generates a flurry of terms at the subsequent order,
i.e.~$\Lambda^{-4}$.
Although we aim at removing all terms with more than one derivative
acting on a field, we find that this does not seem possible.
Therefore, we propose in Proposition \ref{prop:Lambda4reduc1} a field
redefinition that simplifies the most general theory as much as
possible, in the sense that only 4 operators with more than one
derivative acting on a field remain, and all the terms induced by the
field redefinition at order $\Lambda^{-2}$ have been eliminated. 
The theory still contains terms quadratic in double time derivatives
acting on a field, but we show with
Lemma \ref{lemma:kill_double_time_deriv_terms}, that they can be
eliminated with a further field redefinition, leaving the theory with
only linear dependence on double time derivatives on a field.

The linear dependence on double time derivatives, if the term is
having a constant prefactor, is naturally canceled by the Legendre
transform in going to the Hamiltonian.
In fact, it is often considered as a sufficient condition for a
higher-derivative theory to avoid the Ostrogradsky instability or
ghost \cite{Motohashi:2014opa,Motohashi:2016ftl,Crisostomi:2017aim,Motohashi:2017eya,Motohashi:2018pxg,Ganz:2020skf};
in fact, this works out for a single real field or for a real field
theory where the prefactor of the double time derivative contains no
dependence on single time derivatives of other fields and this is the
content of Lemma \ref{lem:avoid_Ostro}.
However, the conjugate momentum picks up new dependence on the double
time derivative of the complex conjugated field at the linear level,
which does not cancel out, and this is because the complex
conjugate of the field is another field and hence the assumption of
the Lemma fails and an Ostrogradsky-like instability is not avoided,
as pointed out in Corollary \ref{coro:fail}.

As a last resort, we choose to relax the assumption of intact Lorentz
invariance, and make a frame dependent low-energy EFT suitable for the
rest frame, which could be considered reasonable for a gapped theory
at low energies.
This is the proposal in Proposition \ref{prop:relax_Lorentz}, where a
suitable field redefinition is found that eliminates all dependence on
double time derivatives in the theory and hence the Ostrogradsky
instability can manifestly be avoided.
The reason why these linear terms in double time derivatives cannot be
eliminated by field redefinitions when insisting on manifestly intact Lorentz
symmetry, like the other problematic terms, is due to an
incompatibility with the metric on $\mathbb{C}P^1$ of the form of said
terms.

Although we have avoided the Ostrogradsky instability at order
$\Lambda^{-4}$ in the derivative expansion of the theory, at the cost
of sacrificing Lorentz invariance, we calculate the corresponding
Hamiltonian of the theory in Lemma \ref{lemma:L4_Hamiltonian} and
prove in Theorem \ref{thm:spiral_instability} that the theory still
suffers from either a spiral instability or a traditional runaway
instability, that is turned on if the norm of the field reaches a
critical value.
Of course, this instability is simply due to the EFT breaking down at
the scale $\Lambda$, but the mathematical nature of the spiral
instability is different from the instability of the EFT breaking down
at the previous ($\Lambda^{-2}$) order in the derivative expansion.

The obvious extensions of our work that one could contemplate, are the
extensions from $\Og(3)$ to $\Og(N)$ or to $\mathbb{C}P^{N-1}$, which
we will leave as future work.
In this direction, it may be interesting to see if it is easier (or
harder) to eliminate higher-order derivative operators in the
$\mathbb{C}P^{N-1}$ case, compared with the $\Og(N)$ case, since the
former enjoys a complex structure and as a target space manifold is
K\"ahler. 
This work has repeatedly utilized integration by parts and discarded
total derivatives, which is sensible on infinite flat Minkowski space
$\mathbb{R}^{d+1}$, but it makes the analysis unsuitable for the
theory if one wishes to apply it to condensed matter physics, like
e.g.~anti-ferromagnetism. In such case, one needs to keep track of
every boundary term and analyze them one-by-one.
We leave such a possibility for future studies.

\subsection*{Acknowledgments}

We thank Lorenzo Bartolini, Johan Bijnens and Chris Halcrow for useful discussions.
S.~B.~G.~thanks the Outstanding Talent Program of Henan University for
partial support.
The work of S.~B.~G.~is supported by the National Natural Science
Foundation of China (Grants No.~11675223 and No.~12071111).
The work of M.N.~is supported in part by JSPS Grant-in-Aid for
Scientific Research (KAKENHI Grant No.~18H01217).

\section{The sigma model}

The $\Og(3)$ or $\CP^1$ \NLSM{} is given by
\beq
\Lag = -\frac12\p_\mu\bn\cdot\p^\mu\bn - \frac\lambda2(\bn\cdot\bn - 1),
\label{eq:L}
\eeq
where $\bn=(n_0,n_1,n_2):\mathbb{R}^{d+1}\to S^2$ is a unit-length 3-vector of real scalar
field and $\lambda$ is a Lagrange multiplier imposing the unit-length
or \NLSM-constraint.
The spacetime index $\mu=0,1,\ldots d$ runs over time and
$d$-dimensional space and we are using the convention in which the
flat Minkowski metric has the mostly-positive signature.

The equation of motion is given by
\beq
\square\bn = \lambda\bn,
\eeq
where $\square=\p_\mu\p^\mu$ is the d'Alembertian operator and by
using the \NLSM-constraint, $\lambda$ is given by
$\bn\cdot\square\bn$. 

The coordinates $\bn$ are called homogeneous coordinates.
Another set of variables natural for the $\Og(N)$ model are the
inhomogeneous coordinates $\bm=(m_1,m_2):\mathbb{R}^{d+1}\to\mathbb{R}^2$, which is an unconstrained
2-vector (or $(N-1)$-vector in the $\Og(N)$ case), related to $\bn$ as
\beq
\bn = \left(\frac{2m_1}{1+\bm\cdot\bm},\frac{2m_2}{1+\bm\cdot\bm},\frac{1-\bm\cdot\bm}{1+\bm\cdot\bm}\right),
\eeq
for which the Lagrangian \eqref{eq:L} is given by
\beq
\Lag = -2\frac{\p_\mu\bm\cdot\p^\mu\bm}{(1+\bm\cdot\bm)^2}.
\label{eq:LO3_inhom}
\eeq
This generalizes to $\Og(N)$ for any $N\geq 2$.

The above Lagrangian is exactly that of the $\CP^1$ model (for $N=3$)
via the identification $\mathbb{C}\ni z=m_1+\i m_2$, for which the
Lagrangian simply reads
\beq
\Lag = -2\frac{\p_\mu z\p^\mu\bar{z}}{(1+|z|^2)^2},
\label{eq:LCP1_inhom}
\eeq
clearly the map from $\bn$ to $z$ is
\beq
\bn = \left(\frac{z+\bar{z}}{1+|z|^2},-\i\frac{z-\bar{z}}{1+|z|^2},\frac{1-|z|^2}{1+|z|^2}\right),
\eeq
and $z$ is the Riemann (2-)sphere coordinate, which is the
stereographic projection from $S^2$ to $\mathbb{C}$.

The advantage of working with inhomogeneous coordinates is that no
constraints need to be taken into account, which will be crucial in
the further analysis in this paper.
Obviously, the Lagrangians \eqref{eq:LO3_inhom} and
\eqref{eq:LCP1_inhom} are identical, but offer straightforward
generalizations to two different theories, namely the $\Og(N)$ \NLSM{}
and the $\CP^{N-1}$ \NLSM, which is the reason for spelling them out
here. 

\subsection{Group invariants and building blocks}

\begin{lemma}
Suppose $\bn$ is a unit-length 3-vector scalar field with
mass-dimension 0, then all $\Og(3)$ and Lorentz invariant terms must
have an even number of derivatives.
\label{lemma:even}
\end{lemma}
\emph{Proof}:
All derivatives contracted by the inverse Minkowski metric come in
pairs, so they must contribute an even number to the total number of
derivatives, say $2n$.
In order to have an odd number of derivatives, we need a Lorentz
invariant tensor structure with an odd number of spacetime indices.
The only one is the Levi-Civita tensor
\beq
\epsilon^{\mu_0\mu_1\cdots\mu_d},
\eeq
which has an odd number of spacetime indices when $d$ is even.
Any combination of derivatives acting on a term that is contracted
with the $\Og(3)$ invariant $\delta_{a b}$ must vanish, due to the
symmetry of the $\Og(3)$ tensor structure versus the anti-symmetry of
the Lorentz invariant Levi-Civita tensor.
The only anti-symmetric $\Og(3)$ invariant tensor is
\beq
\epsilon_{a b c},
\eeq
so one might naively think that the following term is possible
\beq
\epsilon^{\mu\nu\rho}\epsilon_{a b c}\p_\mu n_a\p_\nu n_b\p_\rho n_c.
\eeq
However, geometrically there are only two independent tangent vectors
on $S^2$ and therefore if two derivatives correspond to orthogonal
tangent vectors on $S^2$, the third must be a linear combination of
the latter two and hence an anti-symmetric contraction must vanish.
We have now ruled out any possible terms for $d=2$.
For $d>2$, there are no anti-symmetric group structures of $\Og(3)$ to 
contract with that can give a nonvanishing term.
For $d=0$, there are no nonvanishing terms with one derivative.
This completes the proof.
\hfill$\square$

\subsection{The dimension-2 operators}

\begin{theorem}
Suppose $\bn$ is a unit-length 3-vector scalar field with
mass-dimension 0, then
\beq
\p_\mu\bn\cdot\p^\mu\bn,
\eeq
is the unique dimension 2 term with $\Og(3)$ and Lorentz invariance in
$d+1\neq2$ spacetime dimensions, whereas in $d+1=2$ spacetime
dimensions there is additionally the topological term
\beq
\epsilon^{\mu\nu}\bn\cdot\p_\mu\bn\times\p_\nu\bn.
\eeq
\label{thm:dim2}
\end{theorem}
\emph{Proof}:
Using Lemma \ref{lemma:even}, no terms with a single derivative
exist. 
The $\Og(3)$-invariant tensor structures are $\delta_{a b}$ and
$\epsilon_{a b c}$ and two derivatives must act on the term
constructed with either tensor, since
\begin{align}
  \bn\cdot\p_\mu\bn = 0,
  \label{eq:NLSM_constraint}\\
  \bn\cdot\bn\times\p_\mu\bn = 0,
  \label{eq:folle}
\end{align}
vanish identically.
The former is the \NLSM{} constraint and vanishes because
\beq
\bn\cdot\p_\mu\bn = \frac12\p_\mu(\bn\cdot\bn) = 0,
\eeq
where $\bn\cdot\bn=1$ and the latter vanishes due to anti-symmetry of
the $\Og(3)$-invariant tensor and symmetry of the two $\bn$'s. 
Hence, no composites can be made out of two $\Og(3)$-invariants with a
single spacetime derivative each.

Considering first the tensor $\delta_{a b}$, we can perform an
integration by parts 
\beq
0 = \p_\mu(\bn\cdot\p^\mu\bn) = \p_\mu\bn\cdot\p^\mu\bn + \bn\cdot\square\bn
\label{eq:ibp2}
\eeq
which vanishes identically due to the \NLSM{} constraint
\eqref{eq:NLSM_constraint}.
Since the left-hand side of eq.~\eqref{eq:ibp2} vanishes,
$\bn\cdot\square\bn$ is equal to $-\p_\mu\bn\cdot\p^\mu\bn$ and there are
no other ways of acting with two derivatives on the $\Og(3)$ 
invariant $\delta_{a b}$.

Considering now the second tensor $\epsilon_{a b c}$, a single
derivative vanishes due to eq.~\eqref{eq:folle} and three
anti-symmetrized derivatives vanish as well, see the proof of Lemma
\ref{lemma:even}. The unique nonvanishing contraction with $\Og(3)$
and Lorentz invariance, with dimension less than 4, is thus
\beq
\epsilon^{\mu\nu}\epsilon_{a b c}n_a\p_\mu n_b\p_\nu n_c
=\epsilon^{\mu\nu}\bn\cdot\p_\mu\bn\times\p_\nu\bn,
\label{eq:topological_term}
\eeq
and hence the theorem follows. \phantom{.}\hfill$\square$

\begin{remark}
The topological term \eqref{eq:topological_term} once integrated over
spacetime, measures the topological degree of the mapping from
(one-point compactified) $(1+1)$-dimensional spacetime 
(more precisely an Euclidean two-dimensional space after a Wick rotation) to $S^2$. 
More
commonly used is the Lorentz vector
\beq
Q^\mu = \epsilon^{\mu\nu\rho}\bn\cdot\p_\nu\bn\times\p_\rho\bn,
\eeq
whose time-component, when integrated over space, represents the
static topological degree from (one-point compactified) 2-dimensional
\emph{space} to $S^2$. 
In some literature, this is known as the baby-Skyrme charge and in
other it is known as the vortex charge or vorticity. 
\end{remark}

\begin{lemma}
  The $\Og(3)$ and Lorentz invariant term
  \beq
  \frac{\i\epsilon^{\mu\nu}\p_\mu z\p_\nu\bar{z}}{(1+|z|^2)^2}F(z,\bar{z})
  \eeq
  does not contribute to the equations of motion.
  $F=\overline{F}$ is a real function of the field $z$ and its complex
  conjugate.
  \label{lemma:topo_term}
\end{lemma}
\emph{Proof}:
The Lagrangian 
\beq
\Lag =
-\i\epsilon^{\mu\nu}\p_\mu z\p_\nu\bar{z}F(z,\bar{z})
\eeq
has the corresponding equation of motion for $\bar{z}$:
\begin{align}
  0 &=
  \p_\nu\left(\i\epsilon^{\mu\nu}\p_\mu z F\right)
  -\i\epsilon^{\mu\nu}\p_\mu z\p_\nu\bar{z}\frac{\p F}{\p\bar{z}}\non
  &= \i\epsilon^{\mu\nu}\p_\mu\p_\nu z F
  +\i\epsilon^{\mu\nu}\p_\mu z\p_\nu z\frac{\p F}{\p z}
\end{align}
which is identically zero due to the antisymmetry of $\epsilon^{\mu\nu}$.
Absorbing the metric factor into $F(z,\bar{z})$ completes the proof.
\hfill$\square$

\begin{corollary}
Suppose $\bn$ is a unit-length 3-vector scalar field with
mass-dimension 0, then
\beq
\p_\mu\bn\cdot\p^\mu\bn,
\eeq
is the unique dimension 2 term with $\Og(3)$ and Lorentz invariance
that contributes to the equation of motion.
\label{coro:dim2unique}
\end{corollary}
\emph{Proof}:
Using Theorem \ref{thm:dim2}, the only other term than
$\p_\mu\bn\cdot\p^\mu\bn$ is given by
$\epsilon^{\mu\nu}\bn\cdot\p_\mu\bn\times\p_\nu\bn$ and is a Lorentz
invariant only in $d+1=2$ dimensions.
The latter term is shown not to contribute to the equations of motion,
by setting $F=1$ (or any constant) in Lemma \ref{lemma:topo_term}.
\hfill$\square$

\subsection{The baby-Skyrme term}

The baby-Skyrme term is a special dimension-4 derivative operator,
given by
\beq
(\bn\cdot\p_\mu\bn\times\p_\nu\bn)(\bn\cdot\p^\mu\bn\times\p^\nu\bn).
\label{eq:Skyrme_term}
\eeq
By using the identity
\beq
\epsilon_{a b c}\epsilon_{d e f}
=\delta_{a d}(\delta_{b e}\delta_{c f} - \delta_{b f}\delta_{c e})
-\delta_{a e}(\delta_{b d}\delta_{c f} - \delta_{b f}\delta_{c d})
+\delta_{a f}(\delta_{b d}\delta_{c e} - \delta_{a e}\delta_{c d}),
\label{eq:identity3epsilon}
\eeq
it is easy to see that only the first term is nonvanishing for the
Skyrme term \eqref{eq:Skyrme_term} due to the \NLSM-constraint
\eqref{eq:NLSM_constraint} and hence the Skyrme term obeys the
identity
\begin{align}
(\bn\cdot\p_\mu\bn\times\p_\nu\bn)(\bn\cdot\p^\mu\bn\times\p^\nu\bn)
&=(\p_\mu\bn\times\p_\nu\bn)\cdot(\p^\mu\bn\times\p^\nu\bn) \non
&=(\p_\mu\bn\cdot\p^\mu\bn)^2 - (\p_\mu\bn\cdot\p_\nu\bn)(\p^\mu\bn\cdot\p^\nu\bn).
\end{align}
It is thus clear that the Skyrme term equivalently can be viewed as
the special combination of the two latter terms with relative
coefficient $1$ and $-1$, respectively.
Although the last term in the second line above has a negative
coefficient, it is clear from the left-hand side that this specific
combination is positive semi-definite on a Euclidean manifold. 

It will prove convenient to rewrite the Skyrme term in inhomogeneous
coordinates:
\beq
8\frac{(\p_\mu z\p^\mu\bar{z})(\p_\nu z\p^\nu\bar{z})
  -(\p_\mu z\p_\nu\bar{z})(\p^\mu z\p^\nu\bar{z})}{(1+|z|^2)^4}.
\eeq

\subsection{Further identities}

Let us note that acting with derivatives on the nonlinear sigma model
constraint, $\bn\cdot\bn=1$, yields
\begin{align}
&
\bn\cdot\p_\mu\bn = 0,
\label{eq:nlconstraint1}\\
&
\p_\mu\bn\cdot\p_\nu\bn
  + \bn\cdot\p_\mu\p_\nu\bn = 0,
\label{eq:nlconstraint2}\\
&
\p_\mu\bn\cdot\p_\nu\p_\rho\bn
+\p_\nu\bn\cdot\p_\mu\p_\rho\bn
+\p_\rho\bn\cdot\p_\mu\p_\nu\bn
+\bn\cdot\p_\mu\p_\nu\p_\rho\bn = 0, 
\label{eq:nlconstraint3}
\end{align}
where the first equation is the already well-used identity
\eqref{eq:NLSM_constraint} and the following equations are
generalizations thereof.
The largest tensor structure we will be needing here is a spin-3
tensor (three free Lorentz indices), since the Lorentz-invariant
operator that can be built from such a tensor must have mass dimension
6 or higher and that will be the largest mass dimension we will
consider in this paper.

Contracting two free Lorentz indices with the inverse Minkowski metric
in eqs.~\eqref{eq:nlconstraint2} and \eqref{eq:nlconstraint3} yields
\begin{align}
&
\p_\mu\bn\cdot\p^\mu\bn
+ \bn\cdot\square\bn = 0,
\label{eq:nlconstraint2b}\\
&
2\p_\mu\p_\nu\bn\cdot\p^\mu\bn
  + \p_\nu\bn\cdot\square\bn
  + \bn\cdot\p_\nu\square\bn = 0,
\label{eq:nlconstraint3b}\\
&
2\p_\mu\p_\rho\bn\cdot\p_\nu\p\rho\bn
+\square\bn\cdot\p_\mu\p_\nu\bn
+\p_\mu\bn\cdot\square\p_\nu\bn
+\p_\mu\square\bn\cdot\p_\nu\bn
+2\p_\rho\bn\cdot\p_\mu\p_\nu\p_\rho\bn\non&
+\bn\cdot\p_\mu\p_\nu\square\bn = 0,
\end{align}
whereas the latter identity is obtained by acting with the d'Alembertian
on eq.~\eqref{eq:nlconstraint2}. 

The similar constraint with two contracted pairs of Lorentz indices
is given by
\begin{align}
 &
4\p_\mu\bn\cdot\p^\mu\square\bn
  + 2\p_\mu\p_\nu\bn\cdot\p^\mu\p^\nu\bn
  + \square\bn\cdot\square\bn
  + \bn\cdot\square^2\bn = 0,
\label{eq:nlconstraint4c}
\end{align}
which is obtained by acting with the d'Alembertian on
eq.~\eqref{eq:nlconstraint2b}. 
Two Lorentz contractions yield a minimum mass dimension-4 operator
and the dimension-5 operator would have one free Lorentz index, which
for making at most dimension-6 operators would vanish, since it can
only be contracted with the term of eq.~\eqref{eq:nlconstraint1}. 

The final constraint contains three pairs of Lorentz-contracted indices
\begin{align}
&
12\p_\mu\p_\nu\bn\cdot\p^\mu\p^\nu\square\bn
+6\p_\mu\square\bn\cdot\p^\mu\square\bn
+6\p_\mu\bn\cdot\p^\mu\square^2\bn\non&\qquad
\mathop+4\p_\mu\p_\nu\p_\rho\bn\cdot\p^\mu\p^\nu\p^\rho\bn 
+3\square\bn\cdot\square^2\bn
+\bn\cdot\square^3\bn = 0,
\label{eq:nlconstraint6d}
\end{align}
which is found by acting with the d'Alembertian on
eq.~\eqref{eq:nlconstraint4c}. 
For a Lorentz invariant operator with mass dimension 6, this is the
only identity.

\section{The sigma model as an EFT}

We will now consider the \NLSM{} as an EFT and make a derivative
expansion, but conserving $\Og(3)$ and Lorentz invariance.
The program we will employ here is to use field redefinitions to
eliminate as many derivative operators as possible.
In order to set up the derivative expansion, we will assume that the
\NLSM{} only has one scale $\Lambda$ and hence up to an irrelevant
overall constant factor, and using Corollary \ref{coro:dim2unique}, we
have 
\begin{align}
  \Lag &=
  -\frac12\Lambda^{d-1}\left\{\p_\mu\bn\cdot\p^\mu\bn +
  \mathcal{O}(\Lambda^{-2}) + \lambda(\bn\cdot\bn-1)\right\}\non
  &=-2\Lambda^{d-1}\left\{\frac{\p_\mu z\p^\mu\bar{z}}{(1+|z|^2)^2}
  +\mathcal{O}(\Lambda^{-2})\right\},
\end{align}
for the theory in $(d+1)$-dimensional spacetime and the order
$\Lambda^{-2}$ term represents fourth-order derivative terms, which
thus have to be suppressed by a factor of $\Lambda^2$.
Since we have assumed that there is only one energy scale in the
theory, it must be proportional to $\Lambda$.
$\lambda$ is a Lagrange multiplier enforcing the unit length of the
3-vector field $\bn$ and we have conveniently written the model in
both the vector ($\bn$) and stereographic ($z$) coordinates.

\subsection{Order \texorpdfstring{$\Lambda^{-2}$}{Lambda**-2}}

We will now write down the most general $\Og(3)$ \NLSM{} to
order $\Lambda^{-2}$ in the EFT expansion.
Recalling Lemma \ref{lemma:even}, there are no $\Og(3)$ and Lorentz
invariant terms with an odd number of derivatives and hence no terms
of order $\Lambda^{-1}$, $\Lambda^{-3}$, $\cdots$ and so on.

The complete list of dimension 4 operators is
\begin{align}
(\p_\mu\bn\cdot\p^\mu\bn)(\p_\nu\bn\cdot\p^\nu\bn)\label{eq:opdim4_1},\\
(\p_\mu\bn\cdot\p_\nu\bn)(\p^\mu\bn\cdot\p^\nu\bn)\label{eq:opdim4_2},\\
\bn\cdot\square^2\bn,\label{eq:opdim4_3}\\
\p_\mu\bn\cdot\p^\mu\square\bn,\label{eq:opdim4_4}\\
\square\bn\cdot\square\bn,\label{eq:opdim4_5}\\
\p_\mu\p_\nu\bn\cdot\p^\mu\p^\nu\bn\label{eq:opdim4_6},\\
(\epsilon^{\mu\nu}\bn\cdot\p_\mu\bn\times\p_\nu\bn)(\epsilon^{\rho\sigma}\bn\cdot\p_\rho\bn\times\p_\sigma\bn),\label{eq:opdim4_7}\\
(\epsilon^{\mu\nu}\bn\cdot\p_\mu\bn\times\p_\nu\bn)(\p_\rho\bn\cdot\p^\rho\bn),\label{eq:opdim4_8}
\end{align}
where we have already eliminated possibilities that are related by the
identities \eqref{eq:nlconstraint2} and \eqref{eq:nlconstraint2b} and
the latter two operators (in eqs.~(\ref{eq:opdim4_6}) and (\ref{eq:opdim4_7})) 
are Lorentz invariant only in $d+1=2$
dimensions. 

The identity \eqref{eq:nlconstraint4c} can be used to eliminate one of
the four dimension-4 operators \eqref{eq:opdim4_3}-\eqref{eq:opdim4_6}.
However, integration by parts (IBP) and discarding total derivatives, relates 
the operators \eqref{eq:opdim4_3}-\eqref{eq:opdim4_6}, which can
easily be shown:
\begin{align}
  \p_\mu(\bn\cdot\p^\mu\square\bn) &=
  \p_\mu\bn\cdot\p^\mu\square\bn + \bn\cdot\square^2\bn,\\
  \p_\mu(\p^\mu\bn\cdot\square\bn) &=
  \square\bn\cdot\square\bn +\p_\mu\bn\cdot\p^\mu\square\bn,\\
  \p_\mu(\p_\nu\bn\cdot\p^\mu\p^\nu\bn) &=
  \p_\mu\p_\nu\bn\cdot\p^\mu\p^\nu\bn + \p_\mu\bn\cdot\p^\mu\square\bn,
\end{align}
yielding
\begin{align}
(\p_\mu\bn\cdot\p^\mu\bn)(\p_\nu\bn\cdot\p^\nu\bn),\label{eq:opdim4reduc_1}\\
(\p_\mu\bn\cdot\p_\nu\bn)(\p^\mu\bn\cdot\p^\nu\bn),\label{eq:opdim4reduc_2}\\
\square\bn\cdot\square\bn,\label{eq:opdim4reduc_3}
\end{align}
where the operator \eqref{eq:opdim4_7} has been eliminated from the
above list due to its relation to the existing operators
\eqref{eq:opdim4_1} and \eqref{eq:opdim4_2}.
First using the identity
\beq
\epsilon^{\mu\nu}\epsilon^{\rho\sigma} =
\delta^{\mu\rho}\delta^{\nu\sigma}-\delta^{\mu\sigma}\delta^{\nu\rho},
\eeq
we have
\beq
(\epsilon^{\mu\nu}\bn\cdot\p_\mu\bn\times\p_\nu\bn)(\epsilon^{\rho\sigma}\bn\cdot\p_\rho\bn\times\p_\sigma\bn)
=2(\bn\cdot\p_\mu\bn\times\p_\nu\bn)(\bn\cdot\p^\mu\bn\times\p^\nu\bn),
\eeq
and then using the identity \eqref{eq:identity3epsilon}, we have
\begin{equation}
(\epsilon^{\mu\nu}\bn\cdot\p_\mu\bn\times\p_\nu\bn)(\epsilon^{\rho\sigma}\bn\cdot\p_\rho\bn\times\p_\sigma\bn)
=2(\p_\mu\bn\cdot\p^\mu\bn)(\p_\nu\bn\cdot\p^\nu\bn)
-2(\p_\mu\bn\cdot\p_\nu\bn)(\p^\mu\bn\cdot\p^\nu\bn).
\label{eq:topo_squared}
\end{equation}
The operator \eqref{eq:opdim4_8}, on the other hand, has been
eliminated due to the following considerations.
Consider a field configuration with a non-negative instanton density
$(\epsilon^{\mu\nu}\bn\cdot\p_\mu\bn\times\p_\nu\bn)\geq0$; for such a
configuration, the operator \eqref{eq:opdim4_8} is bounded from below
and perturbations do not destabilize the theory.
Now consider a parity transformation of the latter configuration
$(x^0,x^1)\to(x^0,-x^1)$; this turns instantons into anti-instanton
and anti-instantons into instantons.
Now the operator \eqref{eq:opdim4_8}, however, is non-positive and is
bounded from above.
This has the dire consequence that perturbations of such a field will
destabilize the theory via a runaway instability.
We have thus justified the assumption that the theory should better be
parity invariant, hence eliminating the operator \eqref{eq:opdim4_8}.

One may wonder if the constraint \eqref{eq:nlconstraint4c} can be used to
eliminate the last operator containing boxes,
i.e.~\eqref{eq:opdim4reduc_3}.
Consider thus
\begin{equation}
(b_1+b)(\bn\cdot\square^2\bn)
+(b_2+4b)(\p_\mu\bn\cdot\p^\mu\square\bn)
+(b_3+2b)(\p_\mu\p_\nu\bn\cdot\p^\mu\p^\nu\bn)
+(b_4+b)(\square\bn\cdot\square\bn).
\end{equation}
If we integrate the first three operators by parts, we get
\begin{align}
(b_1-b_2+b_3+b_4)(\square\bn\cdot\square\bn)
+(b_1+b)\p_\mu(\bn\cdot\p^\mu\square\bn)\qquad\qquad\qquad\non
\mathop+(b_2-b_1-b_3-b)\p_\mu(\p^\mu\bn\cdot\square\bn)
+(b_3+2b)\p_\mu(\p_\nu\bn\cdot\p^\mu\p^\nu\bn),
\end{align}
where the terms with coefficient $b$ vanish due to the constraint
\eqref{eq:nlconstraint4c}.
Unfortunately, the remaining (first) term in the above equation, which
is not a total derivative, does not depend on $b$; therefore, no usage
of adjusting $b$ can eliminate the last term.
The complete list of dimension-4 operators with $\Og(3)$, parity and Lorentz
invariance is thus given by
eqs.~\eqref{eq:opdim4reduc_1}-\eqref{eq:opdim4reduc_3}.

\begin{theorem}
The \NLSM{} with $\Og(3)$, parity and Lorentz invariance having
a single mass scale $\Lambda$, written as
\begin{align}
  \Lag &=
  -\frac{\Lambda^{d-1}}{2}\bigg\{(\p_\mu\bn\cdot\p^\mu\bn)
  +\frac{c_4+c_4'}{\Lambda^2}(\p_\mu\bn\cdot\p^\mu\bn)(\p_\nu\bn\cdot\p^\nu\bn)
  -\frac{c_4}{\Lambda^2}(\p_\mu\bn\cdot\p_\nu\bn)(\p^\mu\bn\cdot\p^\nu\bn)\non
  &\phantom{=-\frac12\Lambda^{d-1}\bigg\{\ }
  +\frac{c_{4\square}}{\Lambda^2}(\square\bn\cdot\square\bn)
  +\mathcal{O}(\Lambda^{-4}) + \lambda(\bn\cdot\bn-1)\bigg\},
  \label{eq:NLSM_Lambda2_org}
\end{align}
subject to the constraints $c_4+c_4'\geq 0$, $c_4'\geq 0$,
$c_{4\square}\geq 0$ can, to order $\Lambda^{-2}$, be reduced to 
\begin{align}
  \Lag &=
  -\frac{\Lambda^{d-1}}{2}\bigg\{(\p_\mu\bn\cdot\p^\mu\bn)
  +\frac{c_4+c_4'+c_{4\square}}{\Lambda^2}(\p_\mu\bn\cdot\p^\mu\bn)(\p_\nu\bn\cdot\p^\nu\bn) \non
  &\phantom{=-\frac12\Lambda^{d-1}\bigg\{\ }
  -\frac{c_4}{\Lambda^2}(\p_\mu\bn\cdot\p_\nu\bn)(\p^\mu\bn\cdot\p^\nu\bn)
  +\mathcal{O}(\Lambda^{-4}) + \lambda(\bn\cdot\bn-1)\bigg\},
  \label{eq:NLSM_Lambda2_reduc}
\end{align}
via a field redefinition, where the latter formulation of the theory
contains no d'Alembertian operators.
\label{thm:NLSM_Lambda2}
\end{theorem}
\emph{Proof}:
First we write the theory \eqref{eq:NLSM_Lambda2_org} in terms of
the stereographic coordinate, $z$:
\begin{align}
\Lag &=
-2\Lambda^{d-1}\bigg\{
\frac{\p_\mu z\p^\mu\bar{z}}{(1+|z|^2)^2} 
+\frac{2c_4+4c_4'+4c_{4\square}}{\Lambda^2}
  \frac{(\p_\mu z\p^\mu\bar{z})^2}{(1+|z|^2)^4}
-\frac{2c_4 + 4c_{4\square}}{\Lambda^2}
  \frac{(\p_\mu z\p^\mu z)(\p_\nu\bar{z}\p^\nu\bar{z})}{(1+|z|^2)^4}\non
&\phantom{=-2\Lambda^{d-1}\bigg\{\ }
+\frac{4c_{4\square}}{\Lambda^2}
  \frac{(\p_\mu z\p^\mu z)(\p_\nu \bar{z}\p^\nu\bar{z})}{(1+|z|^2)^3} 
-\frac{2c_{4\square}}{\Lambda^2}
  \frac{(\p_\mu z\p^\mu z)(\bar{z}\square\bar{z})}{(1+|z|^2)^3}
-\frac{2c_{4\square}}{\Lambda^2}
  \frac{(\p_\mu\bar{z}\p^\mu\bar{z})(z\square z)}{(1+|z|^2)^3}\non
&\phantom{=-2\Lambda^{d-1}\bigg\{\ }
+\frac{c_{4\square}}{\Lambda^2}
  \frac{\square z\square\bar{z}}{(1+|z|^2)^2}
  +\mathcal{O}(\Lambda^{-4})
\bigg\}.\label{eq:general_Lambda2_theory_zcoords}
\end{align}
Now considering the following field redefinition
\beq
z\to z + \frac{1}{\Lambda^2}\psi,
\eeq
to order $\Lambda^{-2}$, the terms generated by $\psi$ can only come
from the first term (the kinetic term) in the theory.

It is easily shown that
\begin{align}
\frac{\p_\mu z\p^\mu\bar{z}}{(1+|z|^2)^2} &\to
\frac{\p_\mu z\p^\mu\bar{z}}{(1+|z|^2)^2}
+\frac{1}{\Lambda^2}\left[
  \frac{\p_\mu \psi\p^\mu\bar{z}}{(1+|z|^2)^2}
  +\frac{\p_\mu z\p^\mu\bar\psi}{(1+|z|^2)^2}
  -\frac{2(\p_\mu z\p^\mu\bar{z})(z\bar\psi+\psi\bar{z})}{(1+|z|^2)^3}
  \right]\non&\phantom{\to\ }
+\mathcal{O}(\Lambda^{-4})\non
&=\frac{\p_\mu z\p^\mu\bar{z}}{(1+|z|^2)^2}\non&\phantom{\to\ }
+\frac{1}{\Lambda^2}\left[
  -\frac{\square z\bar{\psi}}{(1+|z|^2)^2}
  -\frac{\square\bar{z}\psi}{(1+|z|^2)^2}
  +\frac{2(\p_\mu z\p^\mu z)\bar{z}\bar{\psi}}{(1+|z|^2)^3}
  +\frac{2(\p_\mu\bar{z}\p^\mu\bar{z})z\psi}{(1+|z|^2)^3}
\right]\non&\phantom{\to\ }
+\mathcal{O}(\Lambda^{-4}).
\end{align}
An educated guess is to take $\psi\propto\square z$, but unfortunately
one can only remove either the term with two boxes or the two terms
with one box; thus another term in $\psi$ is necessary.
Choosing instead
\beq
\psi = \alpha\square z
  + \gamma\frac{(\p_\mu z\p^\mu z)\bar{z}}{1+|z|^2},
\eeq
a straightforward calculation yields
\begin{align}
\Lag &=
-2\Lambda^{d-1}\bigg\{
\frac{\p_\mu z\p^\mu\bar{z}}{(1+|z|^2)^2} 
+\frac{2c_4 + 4c_4' + 4c_{4\square}}{\Lambda^2}
  \frac{(\p_\mu z\p^\mu\bar{z})^2}{(1+|z|^2)^4} \non
&\phantom{=-2\Lambda^{d-1}\bigg\{\ }
-\frac{2c_4 + 4c_{4\square} + 4\gamma}{\Lambda^2}
  \frac{(\p_\mu z\p^\mu z)(\p_\nu\bar{z}\p^\nu\bar{z})}{(1+|z|^2)^4}
+\frac{4c_{4\square} + 4\gamma}{\Lambda^2}
  \frac{(\p_\mu z\p^\mu z)(\p_\nu \bar{z}\p^\nu\bar{z})}{(1+|z|^2)^3} \non
&\phantom{=-2\Lambda^{d-1}\bigg\{\ }
-\frac{2c_{4\square} + \gamma - 2\alpha}{\Lambda^2}
  \frac{(\p_\mu z\p^\mu z)(\bar{z}\square\bar{z})}{(1+|z|^2)^3}
-\frac{2c_{4\square} + \gamma - 2\alpha}{\Lambda^2}
  \frac{(\p_\mu\bar{z}\p^\mu\bar{z})(z\square z)}{(1+|z|^2)^3} \non
&\phantom{=-2\Lambda^{d-1}\bigg\{\ }
+\frac{c_{4\square} - 2\alpha}{\Lambda^2}
  \frac{\square z\square\bar{z}}{(1+|z|^2)^2}
+\mathcal{O}(\Lambda^{-4})
\bigg\}.
\end{align}
Thus setting $\alpha=\frac{1}{2}c_{4\square}$ and
$\gamma=-c_{4\square}$, we get
\begin{align}
\Lag &=
-2\Lambda^{d-1}\bigg\{
\frac{\p_\mu z\p^\mu\bar{z}}{(1+|z|^2)^2} 
+\frac{2c_4 + 4c_4' + 4c_{4\square}}{\Lambda^2}
  \frac{(\p_\mu z\p^\mu\bar{z})(\p_\nu z\p^\nu\bar{z})}{(1+|z|^2)^4}  \non
&\phantom{=-2\Lambda^{d-1}\bigg\{\ }
-\frac{2c_4}{\Lambda^2}
  \frac{(\p_\mu z\p^\mu z)(\p_\nu\bar{z}\p^\nu\bar{z})}{(1+|z|^2)^4}
\bigg\},
\end{align}
which is thus a box-free Lagrangian density.
It is now easy to see, by comparing the coefficients of the above
equation with those of eq.~\eqref{eq:general_Lambda2_theory_zcoords},
that transforming back to vector coordinates $\bn$, yields the theory
\eqref{eq:NLSM_Lambda2_reduc}.
\hfill$\square$

\begin{remark}
Since we have eliminated all boxes from the Lagrangian
to this order, the equation of motion is of second order -- both in
time and space directions.
The theory to this order in $1/\Lambda$ is thus free from the
Ostrogradsky ghost or related instabilities \cite{Woodard:2015zca}, see app.~\ref{app:Ostrogradsky}.
\end{remark}

\begin{remark}
The constraints $c_4+c_4'\geq 0$, $c_4'\geq 0$ and
$c_{4\square}\geq 0$ are due to constraining the static energy density
to be positive definite, which is easier to do by considering the
eigenvalues of the strain tensor $\p_\mu\bn\cdot\p_\nu\bn$.
It is thus easy to see that the $-c_4$ term will give a positive
energy as long as $c_4<c_4+c_4'$ \cite{Gudnason:2017opo}, hence
yielding the second constraint.
\end{remark}

\subsection{Order \texorpdfstring{$\Lambda^{-4}$}{Lambda**-4}}

We will now consider the next order in the derivative or $1/\Lambda$
expansion of the EFT, and thus write down the list of all possible
operators to order $\Lambda^{-4}$ in the EFT expansion, which is:\\
group 1:
\begin{align}
F_{11}&\equiv(\p_\mu\bn\cdot\p^\mu\bn)(\p_\nu\bn\cdot\p^\nu\bn)(\p_\rho\bn\cdot\p^\rho\bn),\\
F_{12}&\equiv(\p_\mu\bn\cdot\p^\nu\bn)(\p_\nu\bn\cdot\p^\mu\bn)(\p_\rho\bn\cdot\p^\rho\bn),\\
F_{13}&\equiv(\p_\mu\bn\cdot\p^\nu\bn)(\p_\nu\bn\cdot\p^\rho\bn)(\p_\rho\bn\cdot\p^\mu\bn),
\end{align}
group 2:
\begin{align}
F_{21}&\equiv(\p_\mu\bn\cdot\p^\mu\bn)(\bn\cdot\square^2\bn),\label{eq:opdim6_21}\\
F_{22}&\equiv(\p_\mu\bn\cdot\p^\mu\bn)(\p_\nu\bn\cdot\p^\nu\square\bn),\\
F_{23}&\equiv(\p_\mu\bn\cdot\p^\mu\bn)(\square\bn\cdot\square\bn),\label{eq:opdim6_23}\\
F_{24}&\equiv(\p_\mu\bn\cdot\p^\mu\bn)(\p_\nu\p_\rho\bn\cdot\p^\nu\p^\rho\bn),\label{eq:opdim6_24}
\end{align}
group 3:
\begin{align}
F_{31}&\equiv(\p_\mu\bn\cdot\p_\nu\bn)(\bn\cdot\p^\mu\p^\nu\square\bn),\\
F_{32}&\equiv(\p_\mu\bn\cdot\p_\nu\bn)(\p^\mu\bn\cdot\p^\nu\square\bn),\\
F_{33}&\equiv(\p_\mu\bn\cdot\p_\nu\bn)(\p_\rho\bn\cdot\p^\mu\p^\nu\p^\rho\bn),\\
F_{34}&\equiv(\p_\mu\bn\cdot\p_\nu\bn)(\p^\mu\p^\nu\bn\cdot\square\bn),\label{eq:opdim6_34}\\
F_{35}&\equiv(\p_\mu\bn\cdot\p_\nu\bn)(\p^\mu\p^\rho\bn\cdot\p^\nu\p^\rho\bn),
\end{align}
group 4:
\begin{align}
F_{41}&\equiv(\p_\mu\bn\cdot\square\bn)(\bn\cdot\p^\mu\square\bn),\\
F_{42}&\equiv(\p_\mu\bn\cdot\square\bn)(\p^\mu\bn\cdot\square\bn),\label{eq:opdim6_42}\\
F_{43}&\equiv(\p_\mu\bn\cdot\square\bn)(\p^\nu\bn\cdot\p^\mu\p^\nu\bn),\\
F_{44}&\equiv(\bn\cdot\p_\mu\square\bn)(\p_\nu\bn\cdot\p^\mu\p^\nu\bn),\\
F_{45}&\equiv(\p_\mu\bn\cdot\p^\mu\p^\nu\bn)(\p^\rho\bn\cdot\p_\nu\p_\rho\bn),\label{eq:opdim6_45}\\
F_{46}&\equiv(\bn\cdot\p_\mu\square\bn)(\bn\cdot\p^\mu\square\bn),
\end{align}
group 5:
\begin{align}
F_{51}&\equiv(\bn\cdot\p_\mu\p_\nu\p_\rho\bn)(\bn\cdot\p^\mu\p^\nu\p^\rho\bn),\\
F_{52}&\equiv(\bn\cdot\p_\mu\p_\nu\p_\rho\bn)(\p^\mu\bn\cdot\p^\nu\p^\rho\bn),\\
F_{53}&\equiv(\p_\mu\bn\cdot\p_\nu\p_\rho\bn)(\p^\mu\bn\cdot\p^\nu\p^\rho\bn),\label{eq:opdim6_53}\\
F_{54}&\equiv(\p_\mu\bn\cdot\p_\nu\p_\rho\bn)(\p^\nu\bn\cdot\p^\mu\p^\rho\bn),\label{eq:opdim6_54}
\end{align}
group 6:
\begin{align}
F_{61}&\equiv\bn\cdot\square^3\bn,\\
F_{62}&\equiv\p_\mu\bn\cdot\p^\mu\square^2\bn,\\
F_{63}&\equiv\p_\mu\p_\nu\bn\cdot\p^\mu\p^\nu\square\bn,\\
F_{64}&\equiv\square\bn\cdot\square^2\bn,\\
F_{65}&\equiv\p_\mu\p_\nu\p_\rho\bn\cdot\p^\mu\p^\nu\p^\rho\bn,\\
F_{66}&\equiv\p_\mu\square\bn\cdot\p^\mu\square\bn,\label{eq:opdim6_66}
\end{align}
group 7:
\begin{align}
F_{71}&\equiv(\epsilon^{\mu\nu}\bn\cdot\p_\mu\bn\times\p_\nu\bn)(\p_\mu\bn\cdot\p^\mu\bn)(\p_\nu\bn\cdot\p^\nu\bn),\\
F_{72}&\equiv(\epsilon^{\mu\nu}\bn\cdot\p_\mu\bn\times\p_\nu\bn)(\p_\mu\bn\cdot\p_\nu\bn)(\p^\mu\bn\cdot\p^\nu\bn),\\
F_{73}&\equiv(\epsilon^{\mu\nu}\bn\cdot\p_\mu\bn\times\p_\nu\bn)(\bn\cdot\square^2\bn),\\
F_{74}&\equiv(\epsilon^{\mu\nu}\bn\cdot\p_\mu\bn\times\p_\nu\bn)(\p_\mu\bn\cdot\p^\mu\square\bn),\\
F_{75}&\equiv(\epsilon^{\mu\nu}\bn\cdot\p_\mu\bn\times\p_\nu\bn)(\square\bn\cdot\square\bn),\\
F_{76}&\equiv(\epsilon^{\mu\nu}\bn\cdot\p_\mu\bn\times\p_\nu\bn)(\p_\mu\p_\nu\bn\cdot\p^\mu\p^\nu\bn),
\end{align}
which is a total of 34 operators and we have already eliminated
operators that are related by eq.~\eqref{eq:nlconstraint2} to the ones
in the above list. We have furthermore eliminated all operators that
are related by the relation \eqref{eq:topo_squared} to the above
ones. 

Performing integrations by parts on operators with four fields
\begingroup
\allowdisplaybreaks
\begin{align}
\p_\mu\left[(\p_\nu\bn\cdot\p^\nu\bn)(\bn\cdot\p^\mu\square\bn)\right]
 &= F_{21} + F_{22} + 2F_{44},\non
\p_\mu\left[(\p_\nu\bn\cdot\p^\nu\bn)(\p^\mu\bn\cdot\square\bn)\right]
 &= F_{22} + F_{23} + 2F_{43},\non
\p_\mu\left[(\p_\nu\bn\cdot\p^\nu\bn)(\p_\rho\bn\cdot\p^\mu\p^\rho\bn)\right]
 &= F_{22} + F_{24} + 2F_{45},\non
\p_\mu\left[(\p^\mu\bn\cdot\p^\nu\bn)(\bn\cdot\p_\nu\square\bn)\right]
 &= F_{31} + F_{32} + F_{41} + F_{44},\non
\p_\mu\left[(\p_\nu\bn\cdot\p_\rho\bn)(\bn\cdot\p^\mu\p^\nu\p^\rho\bn)\right]
 &= F_{31} + F_{33} + 2F_{52},\non
\p_\mu\left[(\p^\mu\bn\cdot\p^\nu\bn)(\p_\nu\bn\cdot\square\bn)\right]
 &= F_{32} + F_{34} + F_{42} + F_{43},\non
\p_\mu\left[(\p_\nu\bn\cdot\p_\rho\bn)(\p^\nu\bn\cdot\p^\mu\p^\rho\bn)\right]
 &= F_{32} + F_{35} + F_{53} + F_{54},\non
\p_\mu\left[(\p_\nu\bn\cdot\p_\rho\bn)(\p^\mu\bn\cdot\p^\nu\p^\rho\bn)\right]
 &= F_{33} + F_{34} + 2F_{54},\non
\p_\mu\left[(\p^\mu\bn\cdot\p^\nu\bn)(\p^\rho\bn\cdot\p_\nu\p_\rho\bn)\right]
 &= F_{33} + F_{35} + F_{43} + F_{45},\non
\p_\mu\left[(\bn\cdot\square\bn)(\bn\cdot\p^\mu\square\bn)\right]
 &= -F_{21} - F_{22} + F_{41} + F_{46},\non
\p_\mu\left[(\bn\cdot\square\bn)(\p^\mu\bn\cdot\square\bn)\right]
 &= -F_{22} - F_{23} + F_{41} + F_{42},\non
\p_\mu\left[(\bn\cdot\square\bn)(\p_\nu\bn\cdot\p^\mu\p^\nu\bn)\right]
 &= -F_{22} - F_{24} + F_{43} + F_{44},\non
\p_\mu\left[(\bn\cdot\p^\mu\p^\nu\bn)(\bn\cdot\p_\nu\square\bn)\right]
 &= -F_{31} - F_{32} + F_{44} + F_{46},\non
\p_\mu\left[(\bn\cdot\p_\nu\p_\rho\bn)(\bn\cdot\p^\mu\p^\nu\p^\rho\bn)\right]
 &= -F_{31} - F_{33} + F_{51} + F_{52},\non
\p_\mu\left[(\bn\cdot\p^\mu\p^\nu\bn)(\p_\nu\bn\cdot\square\bn)\right]
 &= -F_{32} - F_{34} + F_{41} + F_{43},\non
\p_\mu\left[(\bn\cdot\p_\nu\p_\rho\bn)(\p^\nu\bn\cdot\p^\mu\p^\rho\bn)\right]
 &= -F_{32} - F_{35} + F_{52} + F_{54},\non
\p_\mu\left[(\bn\cdot\p_\nu\p_\rho\bn)(\p^\mu\bn\cdot\p^\nu\p^\rho\bn)\right]
 &= -F_{33} - F_{34} + F_{52} + F_{53},\non
\p_\mu\left[(\bn\cdot\p^\mu\p^\nu\bn)(\p^\rho\bn\cdot\p_\nu\p_\rho\bn)\right]
 &= -F_{33} - F_{35} + F_{44} + F_{45},\label{eq:total_derivatives_4fields}
\end{align}
\endgroup
we find 18 total derivatives and hence 18 relations between the
dimension-6 operators with four fields.
There are 19 of such operators
(i.e.~eq.~\eqref{eq:opdim6_21}-\eqref{eq:opdim6_54}) and 18 relations,
so one would naively think that all but one can be eliminated. 
However, some of the relations are dependent on others; in fact the
rank of the vectors in the space of coefficients is only 13, so we can
only eliminate 13 operators, leaving us with 6 independent operators
composed by six derivatives and four fields.
There is quite a lot of ambiguity in which operators to keep and which
to eliminate. Our choice of which to keep is predicated on symmetry
in the derivatives and the preference of having a box instead of two
uncontracted derivatives acting on the field, because of the field
definitions that we have in mind, to be applied shortly.

Performing instead integrations by parts on the operators with two
fields and six derivatives, we have
\begin{align}
\p_\mu\left[\bn\cdot\p^\mu\square^2\bn\right] &= F_{61} + F_{62},\non
\p_\mu\left[\p_\nu\bn\cdot\p^\mu\p^\nu\square\bn\right] &= F_{62} + F_{63},\non
\p_\mu\left[\p^\mu\bn\cdot\square^2\bn\right] &= F_{62} + F_{64},\non
\p_\mu\left[\p_\nu\p_\rho\bn\cdot\p^\mu\p^\nu\p^\rho\bn\right] &= F_{63} + F_{65},\non
\p_\mu\left[\square\bn\cdot\p^\mu\square\bn\right] &= F_{64} + F_{66},
\label{eq:total_derivatives_2fields}
\end{align}
and hence all but one operator can be eliminated. We will choose to
retain $F_{66}$.

\begin{lemma}
The remaining operators after the integration by parts procedure, are
a complete basis of the most general $\Og(3)$, parity and Lorentz
invariant dimension-6 operators, and they are given by:
\beq
F_{11}, F_{12}, F_{13}, F_{23}, F_{24}, F_{34}, F_{42}, F_{45},
F_{53}, F_{66},
\label{eq:Lag4_reduc_ops}
\eeq
of eqs.~\eqref{eq:opdim6_23}, \eqref{eq:opdim6_24},
\eqref{eq:opdim6_34}, \eqref{eq:opdim6_42}, \eqref{eq:opdim6_45},
\eqref{eq:opdim6_53} and \eqref{eq:opdim6_66}, respectively.
\label{lemma:Lambda4_intbyparts}
\end{lemma}
\emph{Proof}:
First we will write the Lagrangian as
\begin{align}
  \Lag^{(4)} =&\
  c_{11}F_{11}
  +c_{12}F_{12}
  +c_{13}F_{13}\non&
  +c_{21}F_{21}
  +c_{22}F_{22}
  +c_{23}F_{23}
  +c_{24}F_{24}\non&
  +c_{31}F_{31}
  +c_{32}F_{32}
  +c_{33}F_{33}
  +c_{34}F_{34}
  +c_{35}F_{35}\non&
  +c_{41}F_{41}
  +c_{42}F_{42}
  +c_{43}F_{43}
  +c_{44}F_{44}
  +c_{45}F_{45}
  +c_{46}F_{46}\non&
  +c_{51}F_{51}
  +c_{52}F_{52}
  +c_{53}F_{53}
  +c_{54}F_{54}\non&
  +c_{61}F_{61}
  +c_{62}F_{62}
  +c_{63}F_{63}
  +c_{64}F_{64}
  +c_{65}F_{65}
  +c_{66}F_{66}.
  \label{eq:Lag4_general}
\end{align}
Using eqs.~\eqref{eq:total_derivatives_4fields} we obtain
\begingroup
\allowdisplaybreaks
\begin{align}
  F_{21} &= -F_{23} + 2 F_{24} + 8 F_{45},\non
  F_{22} &= -F_{24} - 2 F_{45},\non
  F_{31} &= -2 F_{23} + 2 F_{24} + F_{34} + 2 F_{42} + 6 F_{45},\non
  F_{32} &= \frac{F_{23}}{2} - \frac{F_{24}}{2} - F_{34} - F_{42} - F_{45},\non
  F_{33} &= \frac23 F_{23} - \frac23 F_{24} - F_{34} - \frac23 F_{42} - 2 F_{45} + \frac23 F_{53},\non
  F_{35} &= -\frac{F_{23}}{6} + \frac{F_{24}}{6} + F_{34} + \frac23 F_{42} - \frac23 F_{53},\non
  F_{41} &= F_{23} - F_{24} - F_{42} - 2 F_{45},\non
  F_{43} &= -\frac{F_{23}}{2} + \frac{F_{24}}{2} + F_{45},\non
  F_{44} &= \frac{F_{23}}{2} - \frac{F_{24}}{2} - 3 F_{45},\non
  F_{46} &= -2 F_{23} + 2 F_{24} + F_{42} + 8 F_{45},\non
  F_{51} &= -2 F_{23} + 2 F_{24} + 2 F_{42} + 6 F_{45} + F_{53},\non
  F_{52} &= \frac23 F_{23} - \frac23 F_{24} - \frac23 F_{42} - 2 F_{45} - \frac{F_{53}}{6},\non
  F_{54} &= -\frac{F_{23}}{3} + \frac{F_{24}}{3} + \frac{F_{42}}{3} + F_{45} - \frac{F_{53}}{3},
\end{align}
\endgroup
up to total derivatives and using eqs.~\eqref{eq:total_derivatives_2fields}, we get 
\beq
F_{61} = -F_{66},\quad
F_{62} = F_{66},\quad
F_{63} = -F_{66},\quad
F_{64} = -F_{66},\quad
F_{65} = F_{66},
\eeq
up to total derivatives.
Hence, we can write
\begin{align}
  \Lag^{(4)} =&\
  c_{11}F_{11}
  +c_{12}F_{12}
  +c_{13}F_{13}\non&
  +c_{23}'F_{23}
  +c_{24}'F_{24}
  +c_{34}'F_{34}
  +c_{42}'F_{42}
  +c_{45}'F_{45}
  +c_{53}'F_{53}
  +c_{66}'F_{66},
  \label{eq:Lag4_reduc}
\end{align}
with the new coefficients
\begin{align}
  c_{23}' &= c_{23}  - c_{21} - 2 c_{31} + \frac{c_{32}}{2} + \frac23 c_{33} - \frac{c_{35}}{6} + c_{41} - \frac{c_{43}}{2} + \frac{c_{44}}{2} - 2 c_{46} - 2 c_{51} + \frac23 c_{52} - \frac{c_{54}}{3},\non
  c_{24}' &= c_{24} + 2 c_{21} - c_{22} + 2 c_{31} - \frac{c_{32}}{2} - \frac23 c_{33} + \frac{c_{35}}{6} - c_{41} + \frac{c_{43}}{2} - \frac{c_{44}}{2} + 2 c_{46} + 2 c_{51} - \frac23 c_{52}\non
    &\phantom{=\ }+ \frac{c_{54}}{3},\non
  c_{34}' &= c_{34} + c_{31} - c_{32} - c_{33} + c_{35},\non
  c_{42}' &= c_{42} + 2 c_{31} - c_{32} - \frac23 c_{33} + \frac23 c_{35} - c_{41} + c_{46} + 2 c_{51} - \frac23 c_{52} + \frac{c_{54}}{3},\non
  c_{45}' &= c_{45} + 8 c_{21} - 2 c_{22} + 6 c_{31} - c_{32} - 2 c_{33} - 2 c_{41} + c_{43} - 3 c_{44} + 8 c_{46} + 6 c_{51} - 2 c_{52} + c_{54},\non
  c_{53}' &= c_{53} + \frac23 c_{33} - \frac23 c_{35} + c_{51} - \frac{c_{52}}{3} - \frac{c_{54}}{3},\non
  c_{66}' &= c_{66} - c_{61} + c_{62} - c_{63} - c_{64} + c_{65}.
\end{align}
The remaining operators ($F$s) in the Lagrangian \eqref{eq:Lag4_reduc}
are indeed those of eq.~\eqref{eq:Lag4_reduc_ops}.
\hfill$\square$

At this point, it will prove convenient to introduce the following
short-hand notation:
\begingroup
\allowdisplaybreaks
\begin{align}
M &\equiv 1 + |z|^2,\non
K &\equiv \p_\mu z\p^\mu\bar{z},\non
H &\equiv \p_\mu z\p^\mu z,\non
B &\equiv \square z,\non
E &\equiv (\p_\mu\p_\nu z)(\p^\mu\p^\nu z),\non
F &\equiv (\p_\mu\p_\nu z)(\p^\mu\p^\nu \bar{z}),\non
G &\equiv (\p_\mu\bar{z})(\p_\nu\bar{z})(\p^\mu\p^\nu z),\non
I &\equiv (\p_\mu z)(\p_\nu\bar{z})(\p^\mu\p^\nu z),\non
J &\equiv (\p_\mu z)(\p_\nu z)(\p^\mu\p^\nu z),\non
O &\equiv (\p_\mu\p_\nu z)(\p^\nu\p^\rho \bar{z})(\p_\rho\bar{z})(\p^\mu z),\non
P &\equiv (\p_\mu\p_\nu z)(\p^\nu\p^\rho \bar{z})(\p_\rho z)(\p^\mu\bar{z}),\non
Q &\equiv (\p_\mu\p_\nu z)(\p^\nu\p^\rho z)(\p_\rho\bar{z})(\p^\mu\bar{z}),\non
R &\equiv \p_\nu B \p^\nu\bar{z}
  = (\p^\nu\square z)(\p_\nu\bar{z}),\non
S &\equiv \p_\nu B \p^\nu z
  = (\p^\nu\square z)(\p_\nu z),\non
T &\equiv \tfrac12\p_\nu B\p^\nu \ol{H}
  = (\p^\nu\square z)(\p_\nu\p_\rho\bar{z})(\p^\rho\bar{z}),\non
U &\equiv \p_\nu B\p^\nu\ol{B}
  = (\p^\nu\square z)(\p_\nu\square\bar{z}),\non
V &\equiv (\p_\mu z)(\p_\nu z)(\p_\rho\zb)(\p^\mu\p^\nu\p^\rho\zb),\label{eq:shorthand}
\end{align}
\endgroup
where $M$, $K$, $F$, $O$, $P$ and $U$ are real Lorentz scalar
quantities, whereas
$H$, $B$, $E$, $G$, $I$, $J$, $Q$, $R$, $S$, $T$, and $V$ are complex
Lorentz scalars.
According to Theorem \ref{thm:NLSM_Lambda2}, the order
$\Lambda^{-2}$ Lagrangian in the above short-hand notation thus neatly
reads
\beq
\Lag = -2\Lambda^{d-1}\left\{
\frac{K}{M^2}
+ \frac{2c_4+4c_4'+4c_{4\square}}{\Lambda^2}\frac{K^2}{M^4}
- \frac{2c_4}{\Lambda^2}\frac{H\overline{H}}{M^4}
+ \mathcal{O}\left(\Lambda^{-4}\right)
\right\}.
\label{eq:L4zL2redef}
\eeq

The \NLSM{} with $\Og(3)$, parity and Lorentz invariance having
a single mass scale $\Lambda$ to order $\Lambda^{-4}$, written as
\begingroup
\allowdisplaybreaks
\begin{align}
  \Lag &=
  -\frac{\Lambda^{d-1}}{2}\bigg\{(\p_\mu\bn\cdot\p^\mu\bn)
  +\frac{c_4+c_4'}{\Lambda^2}(\p_\mu\bn\cdot\p^\mu\bn)(\p_\nu\bn\cdot\p^\nu\bn)
  -\frac{c_4}{\Lambda^2}(\p_\mu\bn\cdot\p_\nu\bn)(\p^\mu\bn\cdot\p^\nu\bn)\non
  &\phantom{=-\frac12\Lambda^{d-1}\bigg\{\ }
  +\frac{c_{4\square}}{\Lambda^2}(\square\bn\cdot\square\bn)
  +\frac{c_{11}}{\Lambda^4}(\p_\mu\bn\cdot\p^\mu\bn)(\p_\nu\bn\cdot\p^\nu\bn)(\p_\rho\bn\cdot\p^\rho\bn)\non
  &\phantom{=-\frac12\Lambda^{d-1}\bigg\{\ }
  +\frac{c_{12}}{\Lambda^4}(\p_\mu\bn\cdot\p^\nu\bn)(\p_\nu\bn\cdot\p^\mu\bn)(\p_\rho\bn\cdot\p^\rho\bn)\non
  &\phantom{=-\frac12\Lambda^{d-1}\bigg\{\ }
  +\frac{c_{13}}{\Lambda^4}(\p_\mu\bn\cdot\p^\nu\bn)(\p_\nu\bn\cdot\p^\rho\bn)(\p_\rho\bn\cdot\p^\mu\bn)
  +\frac{c_{23}}{\Lambda^4}(\p_\mu\bn\cdot\p^\mu\bn)(\square\bn\cdot\square\bn)\non
  &\phantom{=-\frac12\Lambda^{d-1}\bigg\{\ }
  +\frac{c_{24}}{\Lambda^4}(\p_\mu\bn\cdot\p^\mu\bn)(\p_\nu\p_\rho\bn\cdot\p^\nu\p^\rho\bn)
  +\frac{c_{34}}{\Lambda^4}(\p_\mu\bn\cdot\p_\nu\bn)(\p^\mu\p^\nu\bn\cdot\square\bn)\non
  &\phantom{=-\frac12\Lambda^{d-1}\bigg\{\ }
  +\frac{c_{42}}{\Lambda^4}(\p_\mu\bn\cdot\square\bn)(\p^\mu\bn\cdot\square\bn)
  +\frac{c_{45}}{\Lambda^4}(\p_\mu\bn\cdot\p^\mu\p^\nu\bn)(\p^\rho\bn\cdot\p_\nu\p_\rho\bn)\non
  &\phantom{=-\frac12\Lambda^{d-1}\bigg\{\ }
  +\frac{c_{53}}{\Lambda^4}(\p_\mu\bn\cdot\p_\nu\p_\rho\bn)(\p^\mu\bn\cdot\p^\nu\p^\rho\bn)
  +\frac{c_{66}}{\Lambda^4}(\p_\mu\square\bn\cdot\p^\mu\square\bn)
  +\mathcal{O}(\Lambda^{-6})\non
  &\phantom{=-\frac12\Lambda^{d-1}\bigg\{\ }
  + \lambda(\bn\cdot\bn-1)\bigg\},
  \label{eq:NLSM_Lambda4_org}
\end{align}
\endgroup
subject to the constraints $c_4+c_4'\geq 0$, $c_4'\geq 0$,
$c_{4\square}\geq 0$, $c_{11}\geq 0$, $c_{12}\geq 0$, $c_{13}\geq 0$,
$c_{23}\geq 0$, $c_{24}\geq 0$, $c_{34}\geq 0$, $c_{42}\geq 0$,
$c_{45}\geq 0$, $c_{53}\geq 0$, $c_{66}\geq 0$
can be written in terms of the Riemann sphere coordinate $z$ and its
complex conjugate as
\beq
\Lag = -2\Lambda^{d-1}\left\{
  \Lag^{(0)}
  + \frac{1}{\Lambda^2}\Lag^{(2)}
  + \frac{1}{\Lambda^4}\Lag^{(4)}
  + \mathcal{O}(\Lambda^{-6})
  \right\},
\label{eq:L4z}
\eeq
with the respective orders
\begingroup
\allowdisplaybreaks
\begin{align}
\Lag^{(0)} =&\ \frac{K}{M^2},\label{eq:L4zL0}\\
\Lag^{(2)} =&\ (2c_4+4c_4'+4c_{4\square})\frac{K^2}{M^4}
-(2c_4+4c_{4\square})\frac{H\Hb}{M^4}
-2c_{4\square}\frac{H\bar{z}\Bb}{M^3}
-2c_{4\square}\frac{\Hb zB}{M^3}
+4c_{4\square}\frac{H\Hb}{M^3}\non&
+c_{4\square}\frac{B\Bb}{M^2},\label{eq:L4zL2}\\
\Lag^{(4)} =&\
c_{\{6,5\}}^{K^3}\left[\frac{K^3}{M^{\{6,5\}}}\right]
+c_{\{6,5,4\}}^{K H\Hb}\left[\frac{K H\Hb}{M^{\{6,5,4\}}}\right]
+c_6^{z^2 H\Hb^2} \left[\frac{z^2 H\Hb^2 + \zb^2 H^2\Hb}{M^6}\right]
\non&
+c_6^{\zb^2 K^2 H} \left[\frac{\zb^2 K^2 H + z^2 K^2\Hb}{M^6}\right]
+c_{\{5,4\}}^{\zb K H\Bb} \left[\frac{\zb K H\Bb + z K\Hb B}{M^{\{5,4\}}}\right]
+c_{\{4,3\}}^{K B\Bb} \left[\frac{K B\Bb}{M^{\{4,3\}}}\right]
\non&
-4c_{42} \left[\frac{\zb H\Hb B + z H\Hb\Bb}{M^5}\right]
+c_4^{H\Bb^2} \left[\frac{H\Bb^2 + \Hb B^2}{M^4}\right]
+c_{53} \left[\frac{H\Eb + \Hb E}{M^4}\right]
\non&
+c_4^{K F} \left[\frac{K F}{M^4}\right]
+c_5^{z K G} \left[\frac{z K G + \zb K\Gb}{M^5}\right]
-8c_{66} \left[\frac{\zb H G + z \Hb\Gb}{M^5}\right]
+4c_{66} \left[\frac{B G + \Bb\Gb}{M^4}\right]
\non&
-4c_{45} \left[\frac{\zb K I + z K\Ib}{M^5}\right]
+c_{\{5,4\}}^{z\Hb I} \left[\frac{z\Hb I + \zb H\Ib}{M^{\{5,4\}}}\right]
+c_{\{4,3\}}^{\Bb I} \left[\frac{\Bb I + B\Ib}{M^{\{4,3\}}}\right]
\non&
+c_5^{z H\Jb} \left[\frac{z H\Jb + \zb\Hb J}{M^5}\right]
+16c_{66} \left[\frac{|z|^2 O}{M^4}\right]
+c_4^P \left[\frac{P}{M^4}\right]
+c_4^Q \left[\frac{Q + \Qb}{M^4}\right]
\non&
-4c_{66} \left[\frac{K R + K\Rb}{M^4}\right]
+6c_{66} \left[\frac{\zb^2 H\Rb + z^2\Hb R}{M^4}\right]
-2c_{66} \left[\frac{z\Bb R + \zb B\Rb}{M^3}\right]
\non&
-2c_{66} \left[\frac{H\Sb + \Hb S}{M^4}\right]
-8c_{66} \left[\frac{z T + \zb\Tb}{M^3}\right]
+c_{66} \left[\frac{U}{M^2}\right]
,\label{eq:L4zL4}
\end{align}
with coefficients
\begin{align}
c_6^{K^3} &= 16c_{11} + 8c_{12} + 4c_{13} + 16c_{23} - 8c_{24} + 8c_{34} - 8c_{45} - 8c_{53} - 16c_{66},\non
c_5^{K^3} &= 16c_{24} + 8c_{45} + 8c_{53} + 32c_{66},\non
c_6^{K H\Hb} &= 8c_{12} + 12c_{13} - 16c_{23} + 8c_{24} - 8c_{34} - 8c_{42} + 16c_{66},\non
c_5^{K H\Hb} &= 16c_{23} + 16c_{34} + 8c_{42} - 32c_{66},\non
c_4^{K H\Hb} &= 36c_{66},\non
c_6^{z^2 H\Hb^2} &= 4c_{42} + 4c_{53} - 8c_{66},\non
c_6^{\zb^2 K^2 H} &= 4c_{45} + 8c_{66},\non
c_5^{\zb K H\Bb} &= -8c_{23} - 4c_{34} - 4c_{42},\non
c_4^{\zb K H\Bb} &= -12c_{66}\non
c_4^{K B\Bb} &= 4c_{23} + 2c_{42} - 2c_{66},\non
c_3^{K B\Bb} &= 4c_{66},\non
c_4^{H\Bb^2} &= c_{42} + c_{66},\non
c_4^{K F} &= 4c_{24} + 2c_{53},\non
c_5^{z K G} &= -8c_{24} - 4c_{45} - 4c_{53} - 16c_{66},\non
c_5^{z\Hb I} &= -4c_{34} + 16c_{66},\non
c_4^{z\Hb I} &= -24c_{66},\non
c_4^{\Bb I} &= 2c_{34} - 4c_{66},\non
c_3^{\Bb I} &= 8c_{66},\non
c_5^{z H\Jb} &= -4c_{53} + 8c_{66},\non
c_4^P &= 2c_{45} + 8c_{66},\non
c_4^Q &= c_{45} + 4c_{66},
\end{align}
\endgroup
where the short-hand notation \eqref{eq:shorthand} has been used and
we have defined the notation
\beq
c_{\{a,b,c\}}\frac{X}{M^{\{a,b,c\}}} = \sum_{n\in\{a,b,c\}}c_n\frac{X}{M^n}.
\eeq

We will now perform field redefinitions of the field $z$ up to order
$\Lambda^{-4}$:
\beq
z \to z + \frac{1}{\Lambda^2}\psi_1 + \frac{1}{\Lambda^4}\psi_2,
\label{eq:z_redef4}
\eeq
where $\psi_1$ was already determined in Theorem
\ref{thm:NLSM_Lambda2} and in the short-hand notation
\eqref{eq:shorthand} is given by
\beq
\psi_1 = c_{4\square}\left(\frac12 B - \frac{H\bar{z}}{M}\right),
\label{eq:L4psi1}
\eeq
whereas $\psi_2$ is only determined at the higher order, namely at
order $\Lambda^{-4}$.
The zeroth order Lagrangian is unchanged under any field redefinition
(by definition), whereas the proof of Theorem \ref{thm:NLSM_Lambda2}
can be restated in the short-hand notation \eqref{eq:shorthand} as
\begin{align}
\delta\Lag^{(2)} &= \frac{\p\Lag^{(0)}}{\p z} \psi_1
+ \frac{\p\Lag^{(0)}}{\p\bar{z}} \bar{\psi}_1
- \p_\mu\left(\frac{\p\Lag^{(0)}}{\p\p_\mu z}\right) \psi_1
- \p_\mu\left(\frac{\p\Lag^{(0)}}{\p\p_\mu\bar{z}}\right) \bar{\psi}_1\non
&=\left(-\frac{\Bb}{M^2} + \frac{2\Hb z}{M^3}\right)\psi_1
+\left(-\frac{B}{M^2} + \frac{2H \zb}{M^3}\right)\psib_1,
\end{align}
and hence we obtain nicely
\beq
\Lag^{(2)} + \delta\Lag^{(2)} =
  (2c_4+4c_4'+4c_{4\square})\frac{K^2}{M^4}
  - 2c_4\frac{H\Hb}{M^4},
\label{eq:L4zL2redef2}
\eeq
which indeed is box free.

The change of the fourth-order Lagrangian due to the field
redefinition \eqref{eq:z_redef4} is given by
\begin{align}
\delta\Lag^{(4)} &= \frac{\p\Lag^{(0)}}{\p z} \psi_2
+ \frac{\p\Lag^{(0)}}{\p\bar{z}} \bar{\psi}_2
+ \frac{\p\Lag^{(0)}}{\p\p_\mu z} \p_\mu\psi_2
+ \frac{\p\Lag^{(0)}}{\p\p_\mu\bar{z}} \p_\mu\bar{\psi}_2 \non
&
+ \frac{1}{2}\frac{\p^2\Lag^{(0)}}{\p z^2}\psi_1^2
+ \frac{1}{2}\frac{\p^2\Lag^{(0)}}{\p\bar{z}^2}\bar{\psi}_1^2
+ \frac{\p^2\Lag^{(0)}}{\p z\p\bar{z}}|\psi_1|^2
+ \frac{\p^2\Lag^{(0)}}{\p\p_\mu z\p\p_\nu\bar{z}}\p_\mu\psi_1\p_\nu\bar{\psi}_1 \non
&
+ \frac{\p^2\Lag^{(0)}}{\p\p_\mu z\p z}\psi_1\p_\mu\psi_1
+ \frac{\p^2\Lag^{(0)}}{\p\p_\mu z\p\bar{z}}\bar{\psi_1}\p_\mu\psi_1
+ \frac{\p^2\Lag^{(0)}}{\p\p_\mu\bar{z}\p z}\psi_1\p_\mu\bar{\psi}_1
+ \frac{\p^2\Lag^{(0)}}{\p\p_\mu\bar{z}\p\bar{z}}\bar{\psi}_1\p_\mu\bar{\psi}_1 \non
&
+ \frac{\p\Lag^{(2)}}{\p z} \psi_1
+ \frac{\p\Lag^{(2)}}{\p\bar{z}} \bar{\psi}_1
+ \frac{\p\Lag^{(2)}}{\p\p_\mu z} \p_\mu\psi_1
+ \frac{\p\Lag^{(2)}}{\p\p_\mu\bar{z}} \p_\mu\bar{\psi}_1
+ \frac{\p\Lag^{(2)}}{\p\square z} \square\psi_1
+ \frac{\p\Lag^{(2)}}{\p\square\bar{z}} \square\bar{\psi}_1.
\end{align}
\begin{remark}
The corrections to $\Lag^{(2)}$ due to field redefinitions are
determined only by $\psi_1$.
$\psi_1$ in turn will affect $\Lag^{(4)}$, but since we fixed $\psi_1$
at the previous order, the impact at order $\Lambda^{-4}$ and hence on
$\Lag^{(4)}$ is given.
$\psi_2$ does not affect $\Lag^{(2)}$ and should thus be used for
simplifying $\Lag^{(4)}$. The perturbative ordering in the field
redefinitions is thus evident.
\end{remark}

The \NLSM{} with $\Og(3)$, parity and Lorentz invariance having
a single mass scale $\Lambda$ to order $\Lambda^{-4}$ \eqref{eq:L4z}
with \eqref{eq:L4zL0}-\eqref{eq:L4zL4} and with the field redefinition
\eqref{eq:z_redef4} and $\psi_1$ of eq.~\eqref{eq:L4psi1} is given by
Lagrangian \eqref{eq:L4z} with the zeroth order unchanged
(eq.~\eqref{eq:L4zL0}), the second order in $1/\Lambda$ by
eq.~\eqref{eq:L4zL2redef2} and the fourth order in $1/\Lambda$ by
\begingroup
\allowdisplaybreaks
\begin{align}
&\Lag^{(4)} + \delta\Lag^{(4)} = 
c_{\{6,5\}}^{K^3} \left[\frac{K^3}{M^{\{6,5\}}}\right]
+c_{\{6,5,4\}}^{K H\Hb} \left[\frac{K H\Hb}{M^{\{6,5,4\}}}\right]
+5c_{4\square}^2 \left[\frac{z^4 K\Hb^2 + \zb^4 K H^2}{M^6}\right]
\non&
+c_{\{6,5\}}^{z^2 H\Hb^2} \left[\frac{z^2 H\Hb^2 + \zb^2\Hb H^2}{M^{\{6,5\}}}\right]
+c_6^{\zb^2 K^2 H} \left[\frac{\zb^2 K^2 H + z^2 K^2\Hb}{M^6}\right]
\non&
+c_{\{5,4\}}^{\zb K H\Bb} \left[\frac{\zb K H\Bb + z K\Hb B}{M^{\{5,4\}}}\right]
-4c_{4\square}^2 \left[\frac{z^3 K\Hb\Bb + \zb^3 K H B}{M^5}\right]
+c_{\{4,3\}}^{K B\Bb} \left[\frac{K B\Bb}{M^{\{4,3\}}}\right]
\non&
+\frac34c_{4\square}^2 \left[\frac{z^2 K\Bb^2 + \zb^2 K B^2}{M^4}\right]
+c_5^{z K^2\Bb} \left[\frac{z K^2\Bb + \zb K^2 B}{M^5}\right]
+c_{\{5,4\}}^{\zb H\Hb B} \left[\frac{\zb H\Hb B + z H\Hb\Bb}{M^{\{5,4\}}}\right]
\non&
+c_{\{4,3\}}^{H\Bb^2} \left[\frac{H\Bb^2 + \Hb B^2}{M^{\{4,3\}}}\right]
-c_{4\square}^2 \left[\frac{z B\Bb^2 + \zb\Bb B^2}{M^3}\right]
+5c_{4\square}^2 \left[\frac{z^2\Hb B\Bb + \zb^2 H B\Bb}{M^4}\right]
\non&
-8c_{4\square}^2 \left[\frac{z^3\Hb^2 B + \zb^3 H^2\Bb}{M^5}\right]
+c_{53} \left[\frac{H\Eb + \Hb E}{M^4}\right]
+c_4^{K F} \frac{K F}{M^4}
+c_5^{z K G} \left[\frac{z K G + \zb K\Gb}{M^5}\right]
\non&
-8c_{66} \left[\frac{\zb H G + z\Hb\Gb}{M^5}\right]
+4c_{66} \left[\frac{B G + \Bb\Gb}{M^4}\right]
+c_5^{\zb K I} \left[\frac{\zb K I + z K\Ib}{M^5}\right]
+c_{\{5,4\}}^{z\Hb I} \left[\frac{z\Hb I + \zb H\Ib}{M^{\{5,4\}}}\right]
\non&
+c_{\{4,3\}}^{\Bb I} \left[\frac{\Bb I + B\Ib}{M^{\{4,3\}}}\right]
+2c_{4\square}^2 \left[\frac{z^2\Bb\Ib + \zb^2 B I}{M^4}\right]
-4c_{4\square}^2 \left[\frac{z^3 \Hb\Ib}{M^5} + \frac{\zb^3 H I}{M^5}\right]
\non&
+c_{\{5,4\}}^{z H\Jb} \left[\frac{z H\Jb + \zb\Hb J}{M^{\{5,4\}}}\right]
+4c_{4\square}^2 \left[\frac{z^2\Jb B + \zb^2 J\Bb}{M^4}\right]
+c_4^{|z|^2O} \left[\frac{|z|^2 O}{M^4}\right]
+c_4^P \left[\frac{P}{M^4}\right]
\non&
+c_4^Q \left[\frac{Q + \Qb}{M^4}\right]
+c_4^{K R} \left[\frac{K R + K\Rb}{M^4}\right]
+c_4^{\zb^2H\Rb} \left[\frac{\zb^2 H\Rb + z^2\Hb R}{M^4}\right]
+c_{4\square}^2 \left[\frac{z^2\Hb\Rb + \zb^2 H R}{M^4}\right]
\non&
+c_3^{z\Bb R} \left[\frac{z\Bb R + \zb B\Rb}{M^3}\right]
-\frac12c_{4\square}^2 \left[\frac{z\Bb\Rb + \zb B R}{M^3}\right]
+c_{\{4,3\}}^{H\Sb} \left[\frac{H\Sb + \Hb S}{M^{\{4,3\}}}\right]
-c_{4\square}^2 \left[\frac{z B\Sb + \zb\Bb S}{M^3}\right]
\non&
+c_3^{z T} \left[\frac{z T + \zb\Tb}{M^3}\right]
+c_2^U \left[\frac{U}{M^2}\right]
\non&
-\left[\frac{\Bb}{M^2} - \frac{2\Hb z}{M^3}\right]\psi_2^\alpha
-\left[\frac{B}{M^2} - \frac{2H\zb}{M^3}\right]\psib_2^\alpha
-\left[\frac{2K\zb}{M^3} - \frac{\p_\mu\zb}{M^2}\p^\mu\right]\psi_2^\beta
-\left[\frac{2K z}{M^3} - \frac{\p_\mu z}{M^2}\p^\mu\right]\psib_2^\beta
\non&
-\left[
  \frac{2z\p^\mu\Hb}{M^3}
  -\frac{\p^\mu\Bb}{M^2}
  +\frac{2}{M^3}\left(\frac{3\Hb}{M} - 2\Hb + \zb\Bb\right)\p^\mu z
  +\frac{2z}{M^3}\left(\Bb - \frac{3z\Hb}{M}\right)\p^\mu\zb
  \right]\p_\mu\psi_2^\eta
\non&
-\left[
  \frac{2\zb\p^\mu H}{M^3}
  -\frac{\p^\mu B}{M^2}
  +\frac{2}{M^3}\left(\frac{3H}{M} - 2H + z B\right)\p^\mu\zb
  +\frac{2\zb}{M^3}\left(B - \frac{3\zb H}{M}\right)\p^\mu z
  \right]\p_\mu\psib_2^\eta
,
\label{eq:L4zL4redef_psi2}
\end{align}
\endgroup
where we have defined the coefficients
\begingroup
\allowdisplaybreaks
\begin{align}
  c_6^{K^3} &= 16c_{11} + 8c_{12} + 4c_{13} + 16c_{23} - 8c_{24} + 8c_{34} - 8c_{45} - 8c_{53} - 16c_{66},\non
  c_5^{K^3} &= 16c_{24} + 8c_{45} + 8c_{53} + 32c_{66},\non
  c_6^{K H\Hb} &= 8c_{12} + 12c_{13} - 16c_{23} + 8c_{24} - 8c_{34} - 8c_{42} + 16c_{66} - 12c_{4\square}^2 - 16c_4'c_{4\square},\non
  c_5^{K H\Hb} &= 16c_{23} + 16c_{34} + 8c_{42} - 32c_{66} - 4c_{4\square}^2,\non
  c_4^{K H\Hb} &= 36c_{66} - 3c_{4\square}^2,\non
  c_6^{z^2H\Hb^2} &= 4c_{42} + 4c_{53} - 8c_{66} - 17c_{4\square}^2 - 12c_4c_{4\square},\non
  c_5^{z^2H\Hb^2} &= 16c_{4\square}^2,\non
  c_6^{\zb^2 K^2 H} &= 4c_{45} + 8c_{66} + 24c_{4\square}^2 + 12c_4c_{4\square} + 24c_4'c_{4\square},\non
  c_5^{\zb K H\Bb} &= -8c_{23} - 4c_{34} - 4c_{42} + 4c_{4\square}^2,\non
  c_4^{\zb K H\Bb} &= -12c_{66} - c_{4\square}^2,\non
  c_4^{K B\Bb} &= 4c_{23} + 2c_{42} - 2c_{66} - \frac32c_{4\square}^2,\non
  c_3^{K B\Bb} &= 4c_{66} + c_{4\square}^2,\non
  c_5^{z K^2\Bb} &= -8c_{4\square}^2 - 4c_4c_{4\square} - 8c_4'c_{4\square},\non
  c_5^{\zb H\Hb B} &= -4c_{42} + 14c_{4\square}^2 + 4c_4c_{4\square},\non
  c_4^{\zb H\Hb B} &= -10c_{4\square}^2,\non
  c_4^{H\Bb^2} &= c_{42} + c_{66} - 3c_{4\square}^2,\non
  c_3^{H\Bb^2} &= 2c_{4\square}^2,\non
  c_4^{K F} & = 4c_{24} + 2c_{53},\non
  c_5^{z K G} &= -8c_{24} - 4c_{45} - 4c_{53} - 16c_{66},\non
  c_5^{\zb K I} &= -4c_{45} - 16c_{4\square}^2 - 8c_4c_{4\square} - 16c_4'c_{4\square},\non  
  c_5^{z\Hb I} &= -4c_{34} + 16c_{66} - 10c_{4\square}^2,\non
  c_4^{z\Hb I} &= -24c_{66} + 10c_{4\square}^2,\non
  c_4^{\Bb I} &= 2c_{34} - 4c_{66} + 2c_{4\square}^2,\non
  c_3^{\Bb I} &= 8c_{66} - 2c_{4\square}^2,\non
  c_5^{z H\Jb} &= -4c_{53} + 8c_{66} + 2c_{4\square}^2 + 8c_4c_{4\square},\non
  c_4^{z H\Jb} &= -8c_{4\square}^2,\non
  c_4^{|z|^2O} &= 16c_{66} - 12c_{4\square}^2,\non
  c_4^P &= 2c_{45} + 8c_{66},\non
  c_4^Q &= c_{45} + 4c_{66},\non
  c_4^{K R} &= -4c_{66} + 4c_{4\square}^2 + 2c_4c_{4\square} + 4c_4'c_{4\square},\non
  c_4^{\zb^2 H\Rb} &= 6c_{66} - \frac52c_{4\square}^2,\non
  c_3^{z\Bb R} &= -2c_{66} + \frac12c_{4\square}^2,\non
  c_4^{H\Sb} &= -2c_{66} - \frac12c_{4\square}^2 - 2c_4c_{4\square},\non
  c_3^{H\Sb} &= 2c_{4\square}^2,\non
  c_3^{z T} &= -8c_{66} + 6c_{4\square}^2,\non
  c_2^U &= c_{66} - \frac34c_{4\square}^2,
\end{align}
\endgroup
and we have defined
\beq
\psi_2 = \psi_2^\alpha + \psi_2^\beta + \square\psi_2^\eta,
\label{eq:psi2_decomp}
\eeq
where the factor multiplying $\psi_2^\alpha$ has been integrated by
parts so as to make it a multiplicative factor of $\psi_2^\alpha$,
whereas the factor of $\psi_2^\beta$ has not.
$\psi_2^\eta$ has been integrated by parts one more time than
$\psi_2^\alpha$, which is possible as it is defined with the box.
For a list of the possible terms we will use to construct
$\psi_2^\alpha$ and $\psi_2^\beta$, see app.~\ref{app:psi2}.

\begin{proposition}
The \NLSM{} with $\Og(3)$, parity and Lorentz invariance having
a single mass scale $\Lambda$ to order $\Lambda^{-4}$ \eqref{eq:L4z}
with field redefinition \eqref{eq:z_redef4} and Lagrangian components
\eqref{eq:L4zL0}, \eqref{eq:L4zL2redef2} and \eqref{eq:L4zL4redef_psi2}
being zeroth, second and fourth order in $1/\Lambda$, respectively,
can be reduced to \eqref{eq:L4z} with Lagrangian components
\eqref{eq:L4zL0}, \eqref{eq:L4zL2redef2} and the fourth order in
$1/\Lambda$ given by
\begin{align}
&\Lag^{(4)} + \delta\Lag^{(4)} =
c_{\{6,5\}}^{K^3}\left[\frac{K^3}{M^{\{6,5\}}}\right]
+c_{6}^{K H\Hb} \left[\frac{K H\Hb}{M^{6}}\right]
+c_{6}^{z^2 K^2\Hb} \left[\frac{z^2 K^2\Hb + \zb^2 K^2 H}{M^{6}}\right]
\non&
-2c_{53}\left[\frac{z^2 H\Hb^2 + \zb^2\Hb H^2}{M^6}\right]
\non&
+c_{53}\left[\frac{H\Eb + \Hb E}{M^4}\right]
+c_4^{K F}\left[\frac{K F}{M^4}\right]
-4c_{53}\left[\frac{\zb H G + z\Hb\Gb}{M^5}\right]
+c_5^{z K\Ib}\left[\frac{z K\Ib + \zb K I}{M^5}\right],
\label{eq:L4zreduc}
\end{align}
with coefficients
\begin{align}
  c_6^{K^3} &= 16c_{11} + 8c_{12} + 4c_{13} + 16c_{23} + 32c_{24} + 8c_{34} - 8c_{45} + 12c_{53} - 16c_{66},\non
  c_5^{K^3} &= -16c_{24} + 8c_{45} - 8c_{53} + 32c_{66},\non
  c_6^{K H\Hb} &= 8c_{12} + 12c_{13} + 8c_{24} + 8c_{34} - 4c_{45} - 4c_{53} - 16c_{66},\non
  c_6^{z^2 K^2\Hb} &= -12c_{24} + 6c_{45} + 14c_{53} + 24c_{66},\non
  c_4^{K F} &= 4c_{24} - 2c_{45} + 2c_{53} - 8c_{66},\non
  c_5^{z K\Ib} &= 8c_{24} - 4c_{45} - 4c_{53} - 16c_{66}.
  \label{eq:reduced_coefficients}
\end{align}
\label{prop:Lambda4reduc1}
\end{proposition}
\emph{Proof}:
Consider the field $\psi_2$ with decomposition \eqref{eq:psi2_decomp} given by
\begin{align}
  \psi_2^\alpha =&\ 
  \alpha_{\{2,1\}}^{K B}\frac{K B}{M^{\{2,1\}}}
  +\alpha_{\{3,2\}}^{\zb K H}\frac{\zb K H}{M^{\{3,2\}}}
  +\alpha_{\{2,1\}}^{I}\frac{I}{M^{\{2,1\}}}
  +\alpha_2^{H\Bb}\frac{H\Bb}{M^2}
  +\alpha_{\{3,2\}}^{z H\Hb} \frac{z H\Hb}{M^{\{3,2\}}}
  +\alpha_1^{z R}\frac{z R}{M}
  \non&
  -\frac{c_{4\square}^2}{2}\frac{z\Rb}{M}
  +\alpha_2^{\Gb}\frac{\Gb}{M^2}
  +\alpha_3^{z K^2}\frac{z K^2}{M^3}
  +\alpha_1^{\zb S}\frac{\zb S}{M}
  +\alpha_2^{z^2\Hb B}\frac{z^2\Hb B}{M^2}
  +2 c_{4\square}^2 \frac{z^2\Ib}{M^2}
  +\alpha_2^{\zb^2 J}\frac{\zb^2 J}{M^2}
  \non&
  +\frac34 c_{4\square}^2\frac{z^2 K\Bb}{M^2}
  +\alpha_3^{\zb^3 H^2}\frac{\zb^3 H^2}{M^3}
  -\frac52 c_{4\square}^2 \frac{z^3 K\Hb}{M^3}
  +\alpha_1^{\zb E}\frac{\zb E}{M}
  -c_{4\square}^2\frac{\zb B^2}{M},\non
  \psi_2^\beta &=
  \beta_3^{z K^2}\frac{z K^2}{M^3}
  +\beta_3^{z H\Hb}\frac{z H\Hb}{M^3}
  +\beta_3^{\zb K H}\frac{\zb K H}{M^3}
  +\beta_{\{2,1\}}^{H\Bb}\frac{H\Bb}{M^{\{2,1\}}}
  +\beta_1^{z B\Bb}\frac{z B\Bb}{M}
  +\beta_3^{K B}\frac{K B}{M^3},\non
  \psi_2^\eta &= \left(\frac38c_{4\square}^2 - \frac12 c_{66}\right) z,
\end{align}
with coefficients
\begingroup
\allowdisplaybreaks
\begin{align}
  \alpha_2^{K B} &= 2 c_{23} + c_{42} + \frac38 c_{4\square}^2 - \frac52 c_{66},\non
  \alpha_1^{K B} &= -\frac14c_{4\square}^2 + 3 c_{66},\non
  \alpha_3^{\zb K H} &= -4 c_{23} - 4 c_{34} - 2 c_{42} - 2 c_{45} - 4 c_{4} c_{4\square} + 8 c_{4}' c_{4\square} + \frac{51}{4} c_{4\square}^2 - 21 c_{66},\non
  \alpha_2^{\zb K H} &= \frac34 c_{4\square}^2 - 9 c_{66},\non
  \alpha_2^I &= 2 c_{34} + c_{45} + 2 c_{4} c_{4\square} - 4 c_{4}' c_{4\square} - \frac{13}{2} c_{4\square}^2 + 10 c_{66},\non
  \alpha_1^I &= -\frac12 c_{4\square}^2 + 6 c_{66},\non
  \alpha_2^{H\Bb} &= c_{42} - c_{4\square}^2 + c_{66},\non
  \alpha_3^{z H\Hb} &= -2 c_{42} + 12 c_{4\square}^2 + 2 c_{66},\non
  \alpha_2^{z H\Hb} &= -\frac{37}{4} c_{4\square}^2 - c_{66},\non
  \alpha_1^{z R} &= \frac18 c_{4\square}^2 - \frac32 c_{66},\non
  \alpha_2^{\Gb} &= c_{45} - 2 c_{4} c_{4\square} - 4 c_{4}' c_{4\square} - 4 c_{4\square}^2 + 12 c_{66},\non
  \alpha_3^{z K^2} &= -4 c_{45} + 4 c_{4} c_{4\square} + 8 c_{4}' c_{4\square} + 8 c_{4\square}^2 - 32 c_{66},\non
  \alpha_1^{\zb S} &= -\frac{23}{8} c_{4\square}^2 + \frac52 c_{66},\non
  \alpha_2^{z^2\Hb B} &= \frac{37}{8} c_{4\square}^2 + \frac12 c_{66},\non
  \alpha_2^{\zb^2 J} &= \frac{17}{2} c_{4\square}^2 - 6 c_{66},\non
  \alpha_3^{\zb^3 H^2} &= \frac54 c_{4\square}^2 + c_{66},\non
  \alpha_1^{\zb E} &= -\frac32 c_{4\square}^2 + 2 c_{66},\non
  \beta_3^{z K^2} &= 4 c_{24} + 2 c_{53},\non
  \beta_3^{z H\Hb} &= -4 c_{4} c_{4\square} + \frac32 c_{4\square}^2 + 2 c_{53} - 2 c_{66},\non
  \beta_3^{\zb K H} &= -2 c_{45} + 4 c_{4} c_{4\square} + 8 c_{4}' c_{4\square} + 8 c_{4\square}^2 - 4 c_{53} - 16 c_{66},\non
  \beta_2^{H\Bb} &= 2 c_{4} c_{4\square},\non
  \beta_1^{H\Bb} &= -\frac34 c_{4\square}^2 + c_{66},\non
  \beta_1^{z B\Bb} &= \frac38 c_{4\square}^2 - \frac12 c_{66},\non
  \beta_3^{K B} &= c_{45} - 2 c_{4} c_{4\square} - 4 c_{4}' c_{4\square} - 4 c_{4\square}^2 + 8 c_{66}.
\end{align}
\endgroup
Adding three further total derivatives
\begin{align}
\left(3 c_{4\square}^2 - 4 c_{66}\right)
  \p_\mu\left[|z|^2\frac{H\p_\nu\zb\p^\mu\p^\nu\zb}{M^4}\right]
-\left(c_{45} + 4c_{66}\right)
  \p_\mu\left[\frac{K B \p^\mu\zb}{M^4}\right]\non
+\left(\frac32 c_{4\square}^2 + 2c_{66}\right)
  \p_\mu\left[\frac{z^2\Hb\Bb\p^\mu z}{M^4}\right],
\end{align}
to the fourth order Lagrangian and we arrive at
eq.~\eqref{eq:L4zreduc}.
For an exhaustive list of the possibilities for $\psi_2^\alpha$ and $\psi_2^\beta$ we have considered, see app.~\ref{app:psi2}, whereas for the exhaustive list of possibilities of total derivatives on the Riemann sphere coordinate, see app.~\ref{app:total_deriv}. 
The proof is completed by direct albeit tedious computations.
\hfill$\square$

\begin{lemma}
The \NLSM{} with $\Og(3)$, parity and Lorentz invariance having
a single mass scale $\Lambda$ to order $\Lambda^{-4}$ \eqref{eq:L4z}
with Lagrangian components \eqref{eq:L4zL0}, \eqref{eq:L4zL2redef2} and
\eqref{eq:L4zreduc}, is not box-free, but $(\p_0^2z\p_0^2\zb)$,
$(\p_0^2z)^2$ and $(\p_0^2\zb)^2$ can be removed by a subsequent field
redefinition
\beq
z \to z + \frac{1}{\Lambda^4}
\left[c_{53}\frac{H\Bb}{M^2} + c_4^{K F}\frac{K B}{M^2}\right],
\label{eq:z_defef_subseq}
\eeq
for which the theory becomes
\begin{align}
&\Lag^{(4)} + \delta\Lag^{(4)} =
c_{\{6,5\}}^{K^3}\left[\frac{K^3}{M^{\{6,5\}}}\right]
+c_{6}^{K H\Hb} \left[\frac{K H\Hb}{M^{6}}\right]
+c_{6}^{z^2 K^2\Hb} \left[\frac{z^2 K^2\Hb + \zb^2 K^2 H}{M^{6}}\right]
\non&
-2c_{53}\left[\frac{z^2 H\Hb^2 + \zb^2\Hb H^2}{M^6}\right]
\non&
+c_{53}\left[\frac{H\Eb + \Hb E}{M^4}\right]
+c_4^{K F}\left[\frac{K F}{M^4}\right]
-4c_{53}\left[\frac{\zb H G + z\Hb\Gb}{M^5}\right]
+c_5^{z K\Ib}\left[\frac{z K\Ib + \zb K I}{M^5}\right]
\non&
-c_{53}\left[\frac{H\Bb^2 + \Hb B^2}{M^4}\right]
-c_4^{K F}\left[\frac{K B\Bb}{M^4}\right]
+2c_{53}\left[\frac{\zb H\Hb B + z H\Hb\Bb}{M^5}\right]
\non&
+2c_4^{K F}\left[\frac{\zb K H\Bb + z K\Hb B}{M^5}\right].
\label{eq:L4zreduc2}
\end{align}
\label{lemma:kill_double_time_deriv_terms}
\end{lemma}
\emph{Proof}:
The field redefinition \eqref{eq:z_defef_subseq}, after dropping a
boundary term, contributes
\begin{equation}
-\left(\frac{\Bb}{M^2} - \frac{2\Hb z}{M^3}\right)
\left(c_{53}\frac{H\Bb}{M^2} + c_4^{K F}\frac{K B}{M^2}\right)
-\left(\frac{B}{M^2} - \frac{2H\zb}{M^3}\right)
\left(c_{53}\frac{\Hb B}{M^2} + c_4^{K F}\frac{K\Bb}{M^2}\right),
\end{equation}
to the Lagrangian, which results in eq.~\eqref{eq:L4zreduc2}.
Writing out the time and spatial derivatives of the fourth-order
Lagrangian \eqref{eq:L4zreduc2}, we have
\begingroup
\allowdisplaybreaks
\begin{align}
&\Lag^{(4)} + \delta\Lag^{(4)} =
c_{\{6,5\}}^{K^3}\left[\frac{\big(-|z_0| + |z_i|^2\big)^3}{M^{\{6,5\}}}\right]
\non&  
+c_{6}^{K H\Hb} \left[\frac{\big(-|z_0|^2 + |z_i|^2\big)\big(-z_0^2 + z_j^2\big)\big(-\zb_0^2 + \zb_k^2\big)}{M^{6}}\right]
\non&
+c_{6}^{z^2 K^2\Hb}\left[\frac{z^2\big(-|z_0|^2 + |z_i|^2\big)^2\big(-\zb_0^2 + \zb_j^2\big) + \cc}{M^{6}}\right]
\non&
-2c_{53}\left[\frac{z^2\big(-z_0^2 + z_i^2\big)\big(-\zb_0^2 + \zb_j^2\big)^2 + \cc}{M^6}\right]
\non&
+c_{53}\left[\frac{\big(-z_0^2 + z_k^2\big)\big(-2\zb_{0i}^2 + \zb_{ij}^2 + 2\zb_{00}\zb_{ii} - \zb_{ii}\zb_{jj}\big) + \cc}{M^4}\right]
\non&
+c_4^{K F}\left[\frac{\big(-|z_0|^2 + |z_k|^2\big)\big(-2|z_{0i}|^2 + |z_{ij}|^2 + z_{00}\zb_{ii} + \zb_{00}z_{ii} - z_{ii}\zb_{jj}\big)}{M^4}\right]
\non&
-2c_{53}\left[\frac{\zb\big(-z_0^2 + z_k^2\big)\big(\zb_0^2z_{00} + \zb_i^2z_{00} - 4\zb_0\zb_iz_{0i} + 2\zb_i\zb_jz_{ij} + \zb_0^2z_{ii} - \zb_i^2z_{jj}\big) + \cc}{M^5}\right]
\non&
+c_5^{z K\Ib}\left[\frac{z\big(-|z_0|^2 + |z_k|^2\big)\big(|z_0|^2\zb_{00} - z_0\zb_i\zb_{0i} - z_i\zb_0\zb_{0i} + z_i\zb_j\zb_{ij}\big) + \cc}{M^5}\right]
\non&
+2c_4^{K F}\left[\frac{\zb\big(-|z_0|^2 + |z_i|^2\big)\big(-z_0^2 + z_j^2\big)\big(-\zb_{00} + \zb_{kk}\big) + \cc}{M^5}\right],
\label{eq:L4zreduc2_time}
\end{align}
\endgroup
from which it is clear that all terms proportional to $z_{00}^2$,
$\zb_{00}^2$ and $|z_{00}|^2$ have been canceled out and we have used the
short-hand notation $z_\mu=\p_\mu z$, etc.
\hfill$\square$

\begin{remark}
  Notice that the last five lines of eq.~\eqref{eq:L4zreduc2_time} are linearly
  dependent on either $z_{00}$ or $\zb_{00}$ (and not both of them).
  Technically, the assumption of non-degeneracy
  ($\frac{\p^2\Lag}{\p z_{00}^2}\neq 0$) is violated. The Ostrogradsky theorem
  implying instability is thus not valid, but unfortunately, that is
  not necessarily sufficient for ruling out instability.
  \label{rmk:lin_z00}
\end{remark}

In order to illustrate the problem of a residual instability, let us
consider a sub-Lagrangian of the full theory \eqref{eq:L4zreduc2}, i.e.,
\begin{align}
\Lag &= -2\Lambda^{d-5}c_4^{KF}\frac{K(F-B\Bb)}{M^4}\non
&=-\frac{K(F-B\Bb)}{M^4}\non
&= -\frac{\big(-|z_0|^2 + |z_i|^2\big)\big(-2|z_{0i}|^2 + |z_{ij}|^2 +
  z_{00}\zb_{ii} + \zb_{00}z_{ii} - z_{ii}\zb_{jj}\big)}{M^4},
\end{align}
where on the second line, we have set the overall factor equal to
unity and in the last line written out the time and spatial
derivatives, explicitly.
Defining now
\begin{align}
  \pi_z &= \frac{\p\Lag}{\p z_0} - \p_t\left(\frac{\p\Lag}{\p z_{00}}\right)\non
  &= \frac{\zb_0\big(-2|z_{0i}|^2 + |z_{ij}|^2 + z_{00}\zb_{ii} + \zb_{00}z_{ii} - z_{ii}\zb_{jj}\big)}{M^4}
  + \frac{\big(-z_{00}\zb_0 - z_0\zb_{00} + z_{0i}\zb_i + z_i\zb_{0i}\big)\zb_{jj}}{M^4}\non
  &\phantom{=\ }
  - \frac{4K\zb_{ii}\big(z_0\zb + z\zb_0\big)}{M^5}, \\
  \pi_w &= \frac{\p\Lag}{\p z_{00}} = -\frac{K\zb_{ii}}{M^4},\\
  \pi_{z_i} &= \frac{\p\Lag}{\p z_{0i}} = \frac{2K\zb_{0i}}{M^4},\\
  w &= z_0,\\
  w_i &= z_{0i},
\end{align}
we can write down the Ostrogradsky Hamiltonian
\begin{align}
  \Ham &= \pi_z z_0 + \pib_z\zb_0 + \pi_w z_{00} + \pib_w\zb_{00}
  + \pi_{z_i}z_{0i} + \pib_{z_i}\zb_{0i} - \Lag\non
  &= \frac{|z_0|^2\big(-6|z_{0i}|^2 + |z_{ij}|^2 + z_{00}\zb_{ii} + \zb_{00}z_{ii} - z_{ii}\zb_{jj}\big)}{M^4}\non
  &\phantom{=\ }
  + \frac{\big(-z_0^2\zb_{00} + z_{0i}\zb_i z_0 + z_i\zb_{0i}z_0\big)\zb_{jj}}{M^4}
  + \frac{\big(-\zb_0^2 z_{00} + \zb_{0i} z_i \zb_0 + \zb_i z_{0i}\zb_0\big)z_{jj}}{M^4}\non
  &\phantom{=\ }
  - \frac{4K\zb_{ii}\big(z_0^2\zb + z|z_0|^2\big)}{M^5}
  - \frac{4K z_{ii}\big(\zb_0^2 z + \zb|z_0|^2\big)}{M^5}
  + \frac{|z_k|^2\big(2|z_{0i}|^2 + |z_{ij}|^2 - z_{ii}\zb_{jj}\big)}{M^4},
\end{align}
where the Legendre transform has canceled out the terms linear in
$z_{00}$ and $\zb_{00}$ between $\pi_wz_{00}+\pib_w\zb_{00}$ and $\Lag$.
But, unfortunately, there remains linear dependence on $z_{00}$ and
$\zb_{00}$ inside $\pi_z$ which is not canceled out in the resulting
Hamiltonian.
Their presence in $\pi_z$ is indeed a necessity, since the third-order derivative present
in the Euler-Lagrange equation of motion corresponding to the
Lagrangian, manifests itself in the Hamilton equation as
\beq
\p_t\pi_z = -\frac{\p\Ham}{\p z}.
\eeq
Writing now the Hamiltonian in terms of the phase-space variables $z$,
$w$, $w_i$, $\pi_z$, $\pi_w$, $\pi_{z_i}$ and their complex
conjugates, we have
\begin{align}
  \Ham &= \pi_z w + \pib_z\wb + \frac12\pi_{z_i}w_i + \frac12\pib_{z_i}\wb_i
  + \frac{\big(-|w|^2 + |z_k|^2\big)\big(|z_{ij}|^2 z_{ii}\zb_{jj}\big)}{M^4}.
\end{align}
Allowing for spatial derivatives of the phase-space variables, since
this is a Hamiltonian density for a field theory, we can see that the
Hamiltonian is indeed still linearly dependent on $\pi_z$ and its
complex conjugate, and the Ostrogradsky instability is thus not avoided.

\begin{lemma}
  A sufficient condition for avoiding the Ostrogradsky instability,
  defined by the presence of a linearly dependent conjugate momentum in
  the corresponding Ostrogradsky Hamiltonian, can be stated as follows:
  The conjugate momentum $\pi_w=\frac{\p\Lag}{\p z_{00}}$ cannot contain
  time derivatives of any fields other than $z$.
  \label{lem:avoid_Ostro}
\end{lemma}
\emph{Proof}:
The trouble of generating terms linear in $z_{00}$ and $\zb_{00}$ in
the conjugate momentum $\pi_z$ can be traced back to the fact that
the prefactor of the linear term in $z_{00}$ contains the time
derivative of another field, i.e.~$\zb_0$ and not only $z_0$.
That is, consider 
\beq
\Lag = c (z_0^p z_{00} + \zb_0^p \zb_{00}), \qquad
c\in\mathbb{R},\quad p\in\mathbb{Z}_+,
\eeq
then we have the corresponding conjugate momentum
\beq
\pi_z = \frac{\p\Lag}{\p z_0} -\p_t\left(\frac{\p\Lag}{\p z_{00}}\right)
= cp z_0^{p-1} z_{00} - cp z_0^{p-1} z_{00} = 0.
\eeq
But consider instead another real term
\beq
\Lag = c |z_0|^{2p} (z_{00} + \zb_{00}), \qquad
c\in\mathbb{R},\quad p\in\mathbb{Z}_+,
\eeq
then we have the conjugate momentum
\begin{align}
\pi_z = \frac{\p\Lag}{\p z_0} -\p_t\left(\frac{\p\Lag}{\p z_{00}}\right)
&= cp|z_0|^{2(p-1)}\zb_0 z_{00} - cp|z_0|^{2(p-1)}\zb_0 z_{00} -
cp|z_0|^{2(p-1)}z_0\zb_{00}\non
&= -
cp|z_0|^{2(p-1)}z_0\zb_{00},
\end{align}
and hence we have generated a term linear in $\zb_{00}$ in $\pi_z$
which is not canceled by anything, as the Hamiltonian gets the induced
terms 
\beq
\Ham\supset - cp|z_0|^{2(p-1)}\left(z_0^2\zb_{00} + \zb_0^2 z_{00}\right),
\eeq
not present in the Lagrangian.
Generalizing the proof to letting $\frac{\p\Lag}{\p z_{00}}$ contain
functions of $z$ and $\zb$ does not alter the conclusion and hence the
Lemma follows.
\hfill$\square$

\begin{corollary}
  Using Lemma \ref{lem:avoid_Ostro}, we can see that every term with
  linear dependence on $z_{00}$ in the Lagrangian
  \eqref{eq:L4zreduc2_time} also has dependence on $\zb_0$ and hence
  does not obey the condition of the Lemma to avoid the Ostrogradsky
  instability.
  \label{coro:fail}
\end{corollary}

The problem of canceling off the terms linear in $z_{00}$ and
$\zb_{00}$ in the Lagrangian by using field redefinitions, is due to
the metric on $\mathbb{C}P^1$ that generates always two terms, for
example in the form proportional to $\frac{\Bb}{M^2}$ and
$-\frac{2\Hb z}{M^3}$.
The stringency lies in keeping intact $\Og(3)$ or $\mathbb{C}P^1$
symmetry as well as Lorentz symmetry.
If we allow for relaxing the latter condition, we can make some
progress on the problem.

\begin{proposition}
The \NLSM{} with $\Og(3)$, parity and Lorentz invariance having
a single mass scale $\Lambda$ to order $\Lambda^{-4}$ \eqref{eq:L4z}
with Lagrangian components \eqref{eq:L4zL0}, \eqref{eq:L4zL2redef2} and
\eqref{eq:L4zreduc2}, contains only terms linearly dependent on
$\p_0^2z$ or $\p_0^2\zb$, which can be removed by a subsequent field
redefinition that includes Lorentz non-invariant terms
\begin{equation}
z \to z + \frac{1}{\Lambda^4}
\left[
  -c_{53}\frac{2H\zb_{ii}}{M^2}
  -c_4^{K F}\frac{K z_{ii}}{M^2}
  +2c_{53}\frac{z\Hb(z_0^2 +z_i^2)}{M^3}
  -c_5^{z K\Ib}\frac{z K|z_0|^2}{M^3}
  +2c_4^{K F}\frac{\zb K H}{M^3}
\right],
\label{eq:z_defef_subsubseq}
\end{equation}
for which the theory becomes
\begingroup
\allowdisplaybreaks
\begin{align}
&\Lag^{(4)} + \delta\Lag^{(4)} =
c_{\{6,5\}}^{K^3}\left[\frac{K^3}{M^{\{6,5\}}}\right]
+c_{6}^{K H\Hb} \left[\frac{K H\Hb}{M^{6}}\right]
+c_{6}^{z^2 K^2\Hb} \left[\frac{z^2 K^2\Hb + \zb^2 K^2 H}{M^{6}}\right]
\non&
-2c_{53}\left[\frac{z^2 H\Hb^2 + \zb^2\Hb H^2}{M^6}\right]
+4c_4^{K F}\left[\frac{|z|^2 K H\Hb}{M^6}\right]
\non&
+c_{53}\left[\frac{H\Eb + \Hb E}{M^4}\right]
+c_4^{K F}\left[\frac{K F}{M^4}\right]
-4c_{53}\left[\frac{\zb H G + z\Hb\Gb}{M^5}\right]
+c_5^{z K\Ib}\left[\frac{z K\Ib + \zb K I}{M^5}\right]
\non&
-c_{53}\left[\frac{H\Bb^2 + \Hb B^2}{M^4}\right]
-c_4^{K F}\left[\frac{K B\Bb}{M^4}\right]
+2c_{53}\left[\frac{\zb H\Hb B + z H\Hb\Bb}{M^5}\right]
\non&
+2c_{53}\left[\frac{H\Bb\zb_{ii} + \Hb B z_{ii}}{M^4}\right]
+c_4^{K F}\left[\frac{K\Bb z_{ii} + K B\zb_{ii}}{M^4}\right]
\non&
-2c_{53}\left[\frac{z\Hb\Bb(z_0^2 +z_i^2) + \zb H B(\zb_0^2 + \zb_i^2)}{M^5}\right]
+c_5^{z K\Ib}\left[\frac{z K\Bb|z_0|^2 + \zb K B |z_0|^2}{M^5}\right]
\non&
-4c_{53}\left[\frac{z H\Hb\zb_{ii} + \zb H\Hb z_{ii}}{M^5}\right]
-c_4^{K F}\left[\frac{z K\Hb z_{ii} + \zb K H\zb_{ii}}{M^5}\right]
\non&
+2c_{53}\left[\frac{z^2\Hb^2(z_0^2 +z_i^2) + \zb^2 H^2(\zb_0^2 + \zb_i^2)}{M^6}\right]
-c_5^{z K\Ib}\left[\frac{z^2 K\Hb|z_0|^2 + \zb^2 K H |z_0|^2}{M^6}\right]
.
\label{eq:L4zreduc3_noncovariant}
\end{align}
\endgroup
\label{prop:relax_Lorentz}
\end{proposition}
\emph{Proof}:
The field redefinition \eqref{eq:z_defef_subsubseq}, after dropping 
boundary terms, contributes
\begin{align}
\left(\frac{\Bb}{M^2} - \frac{2\Hb z}{M^3}\right)
\bigg(
  c_{53}\frac{2H\zb_{ii}}{M^2}
  +c_4^{K F}\frac{K z_{ii}}{M^2}
  -2c_{53}\frac{z\Hb(z_0^2 +z_i^2)}{M^3}
  +c_5^{z K\Ib}\frac{z K|z_0|^2}{M^3}
  -2c_4^{K F}\frac{\zb K H}{M^3}
\bigg)\non
+\non
\left(\frac{B}{M^2} - \frac{2H\zb}{M^3}\right)
\bigg(
  c_{53}\frac{2H\zb_{ii}}{M^2}
  +c_4^{K F}\frac{K z_{ii}}{M^2}
  -2c_{53}\frac{z\Hb(z_0^2 +z_i^2)}{M^3}
  +c_5^{z K\Ib}\frac{z K|z_0|^2}{M^3}
  -2c_4^{K F}\frac{\zb K H}{M^3}
\bigg),
\end{align}
to the Lagrangian, which results in eq.~\eqref{eq:L4zreduc3_noncovariant}.
Writing out the time and spatial derivatives of the fourth-order
Lagrangian \eqref{eq:L4zreduc3_noncovariant}, we have
\begingroup
\allowdisplaybreaks
\begin{align}
&\Lag^{(4)} + \delta\Lag^{(4)} =
c_{\{6,5\}}^{K^3}\left[\frac{K^3}{M^{\{6,5\}}}\right]
+c_{6}^{K H\Hb} \left[\frac{K H\Hb}{M^{6}}\right]
+c_{6}^{z^2 K^2\Hb} \left[\frac{z^2 K^2\Hb + \zb^2 K^2 H}{M^{6}}\right]
\non&
-2c_{53}\left[\frac{z^2 H\Hb^2 + \zb^2\Hb H^2}{M^6}\right]
+4c_4^{K F}\left[\frac{|z|^2 K H\Hb}{M^6}\right]
+c_{53}\left[\frac{H\big(-2\zb_{0i}^2 + \zb_{ij}^2 + \zb_{ii}\zb_{jj}\big) + \cc}{M^4}\right]
\non&
+c_4^{K F}\left[\frac{K\big(-2|z_{0i}|^2 + |z_{ij}|^2 + z_{ii}\zb_{jj}\big)}{M^4}\right]
\non&
-2c_{53}\left[\frac{\zb H\big(-4\zb_0\zb_iz_{0i} + 2\zb_i\zb_jz_{ij} + 2\zb_i^2z_{jj}\big) + \cc}{M^5}\right]
\non&
+c_5^{z K\Ib}\left[\frac{z K\big(-z_0\zb_i\zb_{0i} - z_i\zb_0\zb_{0i} + z_i\zb_j\zb_{ij} + |z_0|^2 z_{ii}\big) + \cc}{M^5}\right]
-c_4^{K F}\left[\frac{z K\Hb z_{ii} + \cc}{M^5}\right]
\non&
+2c_{53}\left[\frac{z^2\Hb^2(z_0^2 +z_i^2) + \zb^2 H^2(\zb_0^2 + \zb_i^2)}{M^6}\right]
-c_5^{z K\Ib}\left[\frac{z^2 K\Hb|z_0|^2 + \cc}{M^6}\right].
\label{eq:L4zreduc3_noncovariant_time}
\end{align}
\endgroup
The above Lagrangian contains no $\p_0^2z$ or $\p_0^2\zb$ terms and
hence has a second-order equation of motion and thus no Ostrogradsky instability, but it does not enjoy Lorentz invariance.

\begin{remark}
The Lagrangian \eqref{eq:L4zreduc3_noncovariant_time} is manifestly
free from double time derivatives on any single fields, at the cost of
lost Lorentz invariance.
The spatial part of the Lorentz group $\SO(d)\subset\SO(d,1)$ is
unbroken, however, and a physical interpretation of the theory as a
low-energy EFT (effective field theory), is that the rest frame is a
preferred frame.
\end{remark}

\begin{remark}
Our results are consistent with those of ref.~\cite{Grosse-Knetter:1993tae}, which however utilizes the equations of motion to eliminate the higher-order time derivatives from the EFT. Normally, it is not allowed to modify the Lagrangian using the equations of motion (e.g.~the Dirac Lagrangian would disappear), but it is claimed \cite{Grosse-Knetter:1993tae} that for EFTs, there exists a field redefinition (or canonical transformation) that is equivalent to using the equations of motion, to first order in a given coupling constant. We do not rely on such assumption or perturbativity (apart from the derivative expansion itself), and work directly with field redefinitions. Whether our theory non-manifestly possesses (hidden) Lorentz invariance or not, is an open problem that we will leave for future studies.
\end{remark}

\begin{remark}
Although the Lagrangian \eqref{eq:L4zreduc3_noncovariant_time}
manifestly avoids the Ostrogradsky instability, it is not yet
manifestly free from runaway instabilities due to the nonlinearities
in the first-order time derivatives.
\end{remark}

In order to contemplate the stability of the energy corresponding to
the Lagrangian \eqref{eq:L4zreduc3_noncovariant_time}, we will
calculate its corresponding Hamiltonian.

\begin{lemma}
The \NLSM{} with $\Og(3)$, parity and Lorentz invariance having
a single mass scale $\Lambda$ to order $\Lambda^{-4}$ \eqref{eq:L4z}
with Lagrangian components \eqref{eq:L4zL0}, \eqref{eq:L4zL2redef2} and
\eqref{eq:L4zreduc3_noncovariant}, has a corresponding Hamiltonian
given by
\begin{equation}
\Ham = 2\Lambda^{d-1}\left\{\Ham^{(0)} + \frac{1}{\Lambda^2}\Ham^{(2)} + \frac{1}{\Lambda^{4}}\Ham^{(4)} + \mathcal{O}(\Lambda^{-6})\right\},
\end{equation}
with
\begingroup
\allowdisplaybreaks
\begin{align}
  \Ham^{(0)} &= \frac{\Ktilde}{M^2},\\
  \Ham^{(2)} &= (2c_4 + 4c_4' + 4c_{4\square})\frac{K(3|z_0|^2 + |z_j|^2)}{M^4}
  - 2c_4\frac{-3|z_0|^4 + z_0^2\zb_i^2 + \zb_0^2 z_i^2 + z_i^2\zb_j^2}{M^4},\\
  \Ham^{(4)} &=
  c_{\{6,5\}}^{K^3}\left[\frac{K^2(5|z_0|^2 + |z_i|^2)}{M^{\{6,5\}}}\right]
  +c_{6}^{K H\Hb} \left[\frac{\Ktilde H\Hb + 2K H\zb_0^2 + 2K\Hb z_0^2}{M^{6}}\right]
  \non&
  +c_{6}^{z^2 K^2\Hb} \left[\frac{z^2 K\Hb(3|z_0|^2 + |z_i|^2) + 2z^2 K^2\zb_0^2 + \cc}{M^{6}}\right]
  \non&
  -2c_{53}\left[\frac{z^2\Hb^2\Htilde + 4z^2 H\Hb z_0^2 + \cc}{M^6}\right]
  +4c_4^{K F}\left[\frac{|z|^2(\Ktilde H\Hb + 2K H\zb_0^2 + 2K\Hb z_0^2)}{M^6}\right]
  \non&
  +c_{53}\left[\frac{\Htilde(\zb_{ij}^2 + \zb_{ii}\zb_{jj}) + 2(-3z_0^2 + z_i^2)\zb_{0j}^2 + \cc}{M^4}\right]
  \non&
  +c_4^{K F}\left[\frac{\Ktilde(|z_{ij}|^2 + z_{ii}\zb_{jj}) + 2(-3|z_0|^2 + |z_i|^2)|z_{0j}|^2}{M^4}\right]
  \non&
-2c_{53}\left[\frac{2\zb\Htilde(\zb_i\zb_jz_{ij} + \zb_i^2z_{jj}) + 4\zb(-3z_0^2 + z_i^2)\zb_0\zb_j z_{0j} + \cc}{M^5}\right]  
  \non&
  +c_5^{z K\Ib}\left[\frac{z\Ktilde z_i\zb_j\zb_{ij} - z(-3|z_0|^2 + |z_i|^2)(-z_0\zb_i\zb_{0i} - z_i\zb_0\zb_{0i} + |z_0|^2 z_{ii}) + \cc}{M^5}\right]
  \non&
  -c_4^{K F}\left[\frac{z\Ktilde\Hb z_{ii} + 2z\zb_0^2 K z_{ii} + \cc}{M^5}\right]
  \non&
  +2c_{53}\left[\frac{\zb^2(3z_0^2 + z_i^2) H (\zb_0^2 +\zb_i^2) - 2\zb^2\zb_0^2 H^2 + \cc}{M^6}\right]
  \non&
  -c_5^{z K\Ib}\left[\frac{\zb^2(3|z_0|^2 - |z_i|^2)H|z_0|^2 + 2z_0^2\zb^2 K |z_0|^2 + \cc}{M^6}\right],
  \label{eq:H4z}
\end{align}
\endgroup
and the definitions
\beq
\Ktilde = |z_0|^2 + |z_i|^2, \qquad
\Htilde = z_0^2 + z_i^2, \qquad
\Hbtilde = \zb_0^2 + \zb_i^2.
\eeq
\label{lemma:L4_Hamiltonian}
\end{lemma}
\emph{Proof}:
Defining
\begin{align}
  \pi_z &= \frac{\p\Lag}{\p z_0}
  = -2\Lambda^{d-1}\left\{\pi_z^{(0)} + \frac{1}{\Lambda^2}\pi_z^{(2)} + \frac{1}{\Lambda^4}\pi_z^{(4)} + \mathcal{O}(\Lambda^{-6})\right\},\\
  \pi_{z_i} &= \frac{\p\Lag}{\p z_{0i}}
  =-2\Lambda^{d-1}\left\{\frac{1}{\Lambda^4}\pi_{z_i}^{(4)} + \mathcal{O}(\Lambda^{-6})\right\},
\end{align}
the Hamiltonian is written via the Legendre transform as
\beq
\Ham = \pi_z z_0 + \pib_z\zb_0 + \pi_{z_i}z_{0i} + \pib_{z_i}\zb_{0i} - \Lag,
\eeq
and a long but straightforward calculation yields
\begingroup
\allowdisplaybreaks
\begin{align}
  \pi_z^{(0)} &= -\frac{\zb_0}{M^2},\\
  \pi_z^{(2)} &= -(2c_4 + 4c_4' + 4c_{4\square})\frac{2K\zb_0}{M^4} + 2c_4\frac{2\Hb z_0}{M^4},\\
  \pi_z^{(4)} &=
  -c_{\{6,5\}}^{K^3}\frac{3K^2\zb_0}{M^{\{6,5\}}}
  -c_{6}^{K H\Hb} \frac{(\zb_0 H + 2z_0 K)\Hb}{M^{6}}
  -c_{6}^{z^2 K^2\Hb} \frac{2\zb_0(z^2 K\Hb + \zb^2 K H) + 2z_0\zb^2 K^2}{M^{6}}
  \non&
  +2c_{53}\frac{2z_0(z^2\Hb^2 + 2\zb^2 H\Hb)}{M^6}
  -4c_4^{K F}\frac{|z|^2(\zb_0 H + 2z_0 K)\Hb}{M^6}
  \non&
  -c_{53}\frac{2z_0(-2\zb_{0i}^2 + \zb_{ij}^2 + \zb_{ii}\zb_{jj})}{M^4}
  -c_4^{K F}\frac{\zb_0(-2|z_{0i}|^2 + |z_{ij}|^2 + z_{ii}\zb_{jj})}{M^4}
  \non&
  +2c_{53}\frac{2z_0\zb(-4\zb_0\zb_iz_{0i} + 2\zb_i\zb_jz_{ij} + 2\zb_i^2z_{jj}\big) + 4z\Hb z_i\zb_{0i}}{M^5}
  \non&
  -c_5^{z K\Ib}\frac{\zb_0 z(-z_0\zb_i\zb_{0i} - z_i\zb_0\zb_{0i} + z_i\zb_j\zb_{ij} + |z_0|^2 z_{ii})}{M^5}
  \non&
  -c_5^{z K\Ib}\frac{\zb_0 \zb(-\zb_0 z_i z_{0i} - \zb_i z_0 z_{0i} + \zb_i z_j z_{ij} + |z_0|^2 \zb_{ii}) - z K(-\zb_i\zb_{0i} - \zb_i z_{0i} + \zb_0 z_{ii} + \zb_0\zb_{ii})}{M^5}
  \non&
  +c_4^{K F}\frac{\zb_0 z\Hb z_{ii} + \zb_0\zb H\zb_{ii} + 2z_0\zb K\zb_{ii}}{M^5}
  -2c_{53}\frac{4z_0\zb^2 H(z_0^2 +z_i^2) - 2z^2\Hb^2 z_0}{M^6}
  \non&
  +c_5^{z K\Ib}\frac{\zb_0 z^2\Hb|z_0|^2 + \zb_0\zb^2 H |z_0|^2 - z^2 K\Hb\zb_0 - \zb^2 K H\zb_0}{M^6},\\
  \pi_{z_i}^{(4)} &=
  -c_{53}\frac{4\Hb z_{0i}}{M^4}
  -c_4^{K F}\frac{2K\zb_{0i}}{M^4}
  +2c_{53}\frac{4\zb H\zb_0\zb_i}{M^5}
  -c_5^{z K\Ib}\frac{\zb K(-\zb_0 z_i - \zb_i z_0)}{M^5}.
\end{align}
\endgroup
A direct calculation yields the Hamiltonian given in eq.~\eqref{eq:H4z}.
\hfill$\square$

\begin{theorem}
The \NLSM{} with $\Og(3)$, parity and Lorentz invariance having
a single mass scale $\Lambda$ to order $\Lambda^{-4}$ \eqref{eq:L4z}
with Lagrangian components \eqref{eq:L4zL0}, \eqref{eq:L4zL2redef2} and
\eqref{eq:L4zreduc3_noncovariant}, is not bounded from below in the
limit of large time derivatives, and the theory potentially suffers
from a spiral instability or a runaway instability.
\label{thm:spiral_instability}
\end{theorem}
\emph{Proof}:
We start by taking the limit
\begin{align}
  \lim_{|z_i|\to0}\Ham^{(4)} &=
  A\frac{5|z_0|^6}{M^6}
  +B\frac{5|z_0|^4(z^2\zb_0^2 + \zb^2 z_0^2)}{M^6},
\end{align}
corresponding to large time derivatives (and negligible spatial
derivatives) and we have used that $c_5^{K^3}+4c_4^{K F}=0$ and
defined
\beq
A = c_6^{K^3} + c_6^{K H\Hb} - 4c_4^{K F}, \qquad
B = c_6^{z^2K^2\Hb} - 4c_{53} + c_5^{z K\Ib}.
\eeq
Writing $z=\rho e^{\i\theta}$, we have
\beq
z_0 = (\rho_0 + \i\theta_0\rho) e^{\i\theta},
\eeq
and hence
\begin{align}
  |z_0|^6 &= (\rho_0^2 + \rho^2\theta_0^2)^3,\\
  |z_0|^4(z^2\zb_0^2 + \zb^2 z_0^2) &= (\rho_0^2 + \rho^2\theta_0^2)^2
  2\rho^2(\rho_0^2 - \rho^2\theta_0^2).
\end{align}
We assume that $A>0$ is strictly positive, which is equal to the
condition
\beq
4c_{11} + 4c_{12} + 4c_{13} + 4c_{23} + 6c_{24} + 4c_{34} - c_{45} > 0.
\eeq
The spiral instability can occur if $B>0$ and
\beq
A < 2B\rho^2,
\eeq
since in this case the coefficient of $\theta_0^6$ is negative and a
fast rotating phase can lower the energy indefinitely.

On the other hand, we may tune the theory so as to have $B<0$, for
which a runaway instability can occur if
\beq
A < 2|B|\rho^2.
\eeq
In either case, the critical modulus of the field is
\beq
\rho^{\rm crit} = \left|\frac{A}{2B}\right|
=\left|\frac{4c_{11} + 4c_{12} + 4c_{13} + 4c_{23} + 6c_{24} + 4c_{34}
  - c_{45}}{-2c_{24} + c_{45} + 6c_{53} + 4c_{66}}\right|,
\eeq
which when reached can drive the theory into lower and lower energies.
\hfill$\square$

\begin{remark}
The unveiled instability can be avoided if $B=0$, which however would
require a precise fine-tuning of theory, which is not expected 
generically to be the case.
\end{remark}

\begin{remark}
The instability found in Theorem \ref{thm:spiral_instability} is due
to the nature of the derivative expansion of the effective low-energy
theory and is a classical instability. 
The quantum considerations are beyond the scope of this paper.
The instability of spiral type is mathematically different
from the instability taking place at order $\Lambda^{-2}$, which
however is always of the isotropic type.
\end{remark}

\appendix
\section{Ostrogradsky's theorem}\label{app:Ostrogradsky}

For convenience, we review the Ostrogradsky's theorem, specialized to
second-order derivatives:
\begin{theorem}[Ostrogradsky \cite{Ostrogradsky:1850fid}, Woodard \cite{Woodard:2015zca}] 
Given a Lagrangian theory with qua\-dratic and non-degenerate dependence
on a second-order time derivative of a field, the corresponding
Ostrogradsky Hamiltonian possesses a linear dependence on one of the
two conjugate momenta, and resultantly the corresponding energy is not
bounded neither from below nor from above.
\label{thm:ostrogradsky}
\end{theorem}
\emph{Proof}:
Consider the Lagrangian $L(z,z_0,z_{00})$ which is
a functional of $z$, as well as its first
and second-order time derivatives denoted by one and two zeros as
indices, respectively, and the corresponding Ostrogradsky Hamiltonian
$H(z,w,\pi_z,\pi_w)$ which is a functional of $z$, $w=z_0$ and their
conjugate momenta $\pi_z$ and $\pi_w$.\footnote{Here we will use the
  notation that the functional dependence on $z$ includes the dependence on $\zb$
  and will not state so explicitly in order not to clutter the
  notation.}
The condition that the Lagrangian depends non-degenerately on $z_{00}$
means that
\beq
\frac{\p^2 L}{\p z_{00}^2} \neq 0.
\label{eq:Ostro_nondegeneracy}
\eeq
Then the Hamiltonian is defined as
\beq
H = \pi_z z_0 + \pib_z\zb_0 + \pi_w w_0 + \pib_w\wb_0 - L,
\eeq
the conjugate momenta are given by
\begin{align}
  \pi_z &= \frac{\p L}{\p z_0} - \p_t\left(\frac{\p L}{\p z_{00}}\right),\qquad
  \pi_w = \frac{\p L}{\p z_{00}},
\end{align}
and the Hamilton's equations are
\begin{align}
  \frac{\p H}{\p\pi_z} &= z_0 = w, \qquad &
  \frac{\p H}{\p z} &= -\p_t\pi_z,\label{eq:phasespace_tr1}\\
  \frac{\p H}{\p\pi_w} &= w_0, \qquad &
  \frac{\p H}{\p w} &= -\p_t\pi_w.\label{eq:phasespace_tr2}
\end{align}
Finally, the Hamiltonian can be written as
\beq
H = \pi_z w + \pib_z\wb + \pi_w a(z,w,\pi_w) + \pib_w \overline{a(z,w,\pi_w)}
- L(z,w,a(z,w,\pi_w)),
\label{eq:H_Ostrogradsky}
\eeq
where the acceleration $a(z,w,\pi_w)$ is defined by
\beq
\left.\frac{\p L}{\p z_{00}}\right|_{z_0=w,z_{00}=a} = \pi_w.
\label{eq:indirect_acceleration}
\eeq
The Hamilton equations $z_0=\frac{\p H}{\p\pi_z}$,
$w_0=\frac{\p H}{\p\pi_w}$, $\p_t\pi_w=-\frac{\p H}{\p w}$ simply
reproduce the phase space transformation
\eqref{eq:phasespace_tr1}-\eqref{eq:phasespace_tr2}, whereas
$\p_t\pi_z=-\frac{\p H}{\p z}$ reproduces the Euler-Lagrange equation
of $L$. 
The assumption of non-degeneracy \eqref{eq:Ostro_nondegeneracy}
implies that the phase space transformation
\eqref{eq:phasespace_tr1}-\eqref{eq:phasespace_tr2} can be inverted to
solve for $z_{00}$ in terms of $z$, $w$ and $\pi_w$, by means of the implicit function theorem, which is the
statement \eqref{eq:indirect_acceleration}.
Crucially, the acceleration $a(z,w,\pi_w)$ does not depend on the
conjugate momentum $\pi_z$, which is only needed for the third-order
time derivative of $z$.
Notice that the third-order time derivative only appears when the
assumption of non-degeneracy \eqref{eq:Ostro_nondegeneracy} holds
true.

Finally, we have that the Ostrogradsky Hamiltonian
\eqref{eq:H_Ostrogradsky} depends only linearly on the conjugate
momentum $\pi_z$.
The linear dependence implies the Ostrogradsky instability, since the
system can linearly be driven to lower and lower (or higher and
higher) energies.
This completes the proof.
\hfill$\square$

\subsection{Example}

As a simple example, let us consider an extension of the Harmonic
oscillator as
\beq
L = -\frac12\alpha z_{00}^2 + \frac12\beta z_0^2 - \frac12\gamma z^2,
\eeq
with $z\in\mathbb{R}$ being a real field (or function of time),
$\alpha>0$, $\beta>0$, $\gamma>0$ and the indices denote time
derivatives. 
First we have
\beq
w=z_0,\qquad
\pi_z = \beta z_0 + \alpha z_{000}, \qquad
\pi_w = -\alpha z_{00},
\label{eq:Ostro_example_transformation}
\eeq
whereas the Ostrogradsky Hamiltonian reads
\begin{align}
  H &= \pi_z w + \pi_w w_0 - L \non
  &=\pi_z w - \frac{\pi_w^2}{2\alpha} - \frac{\beta w^2}{2} +
  \frac{\gamma z^2}{2},
\end{align}
where the linear dependence on $\pi_z$ is manifest.
Inserting the explicit expression
\eqref{eq:Ostro_example_transformation} for $\pi_z$
\beq
H = \frac{\beta w^2}{2} + \alpha w z_{000} - \frac{\pi_w^2}{2\alpha} + \frac{\gamma z^2}{2},
\eeq
the kinetic energy for $z$, i.e.~$w^2$ is manifestly positive.
Integrating by parts and dropping the boundary term, we can write
\beq
H = \frac{\beta w^2}{2} - \frac{3\pi_w^2}{2\alpha} + \frac{\gamma z^2}{2},
\eeq
from which the terminology Ostrogradsky's ghost comes from; that is, the
kinetic energy, $\pi_w^2$ is negative semi-definite.

\section{Field redefinition operators}\label{app:psi2}

In order to cancel the unwanted operators, we need a systematic
educated guess for the field redefinitions, $\psi_2$, at the order
$\Lambda^{-4}$ and they read
\begin{align}
  &\frac{z K^2}{M^\#},\
  \frac{\zb K H}{M^\#},\
  \frac{z^3 K\Hb}{M^\#},\ 
  \frac{K B}{M^\#},\
  \frac{z^2 K\Bb}{M^\#},\ 
  \frac{z H\Hb}{M^\#},\
  \frac{\zb^3 H^2}{M^\#},\
  \frac{z^5\Hb^2}{M^\#},\
  \frac{\zb^2 H B}{M^\#},\
  \frac{H\Bb}{M^\#},\
  \frac{z^2\Hb B}{M^\#},\non&
  \frac{z^4\Hb\Bb}{M^\#},\ 
  \frac{z B\Bb}{M^\#},\
  \frac{\zb B^2}{M^\#},\
  \frac{z^3\Bb^2}{M^\#},\
  \frac{\zb E}{M^\#},\
  \frac{z^3\Eb}{M^\#},\
  \frac{z F}{M^\#},\
  \frac{z^2 G}{M^\#},\
  \frac{\Gb}{M^\#},\
  \frac{I}{M^\#},\
  \frac{z^2\Ib}{M^\#},\
  \frac{\zb^2 J}{M^\#},\
  \frac{z^4\Jb^2}{M^\#},\non&
  \frac{z R}{M^\#},\
  \frac{z\Rb}{M^\#},\
  \frac{\zb S}{M^\#},\
  \frac{z^3\Sb}{M^\#},
\end{align}
which is the complete list of four derivative operators with one
factor of $z$ more than factors of $\zb$ (with ``chirality'' +1),
needed for generating sixth-order derivative terms by field
redefinitions according to eq.~\eqref{eq:z_redef4}, where $M^\#$
includes a term for each integer power necessary.

Notice that additional factors of $|z|^2$, $|z|^4$ are already
included in the above list by using that $|z|^2=M-1$ and hence an
operator, e.g.,
\beq
\frac{|z|^2X}{M^n} = -\frac{X}{M^n} + \frac{X}{M^{n-1}},
\eeq
and since all possible powers of $1/M$ are included, such
possibilities are already taken into account.

\section{Total derivatives}\label{app:total_deriv}

In order to cancel unwanted operators, we also need a systematic
educated guess for writing down total derivative operators that can be
added to the Lagrangian; with six derivatives and four fields
\begin{align}
  &\p_\mu\left(\frac{\Hb z^{\mu\nu}z_\nu + \cc}{M^\#}\right),\
  \p_\mu\left(\frac{\zb z^{\mu\nu}z_{\nu\rho}\zb^\rho + \cc}{M^\#}\right),\
  \p_\mu\left(\frac{\zb\Bb z^{\mu\nu}z_\nu + \cc}{M^\#}\right),\non&
  \p_\mu\left(\frac{\zb z^{\mu\nu}\zb_{\nu\rho}z^\rho + \cc}{M^\#}\right),\
  \p_\mu\left(\frac{\zb B z^{\mu\nu}\zb_\nu + \cc}{M^\#}\right),\
  \p_\mu\left(\frac{z\Bb z^{\mu\nu}\zb_\nu + \cc}{M^\#}\right),\non&
  \p_\mu\left(\frac{K z^{\mu\nu}\zb_\nu + \cc}{M^\#}\right),\
  \p_\mu\left(\frac{z z^{\mu\nu}\zb_{\nu\rho}\zb^\rho + \cc}{M^\#}\right),\
  \p_\mu\left(\frac{\Ib z^\mu + \cc}{M^\#}\right),\
  \p_\mu\left(\frac{\Hb B z^\mu + \cc}{M^\#}\right),\non&
  \p_\mu\left(\frac{z\Bb^2 z^\mu + \cc}{M^\#}\right),\
  \p_\mu\left(\frac{\zb B\Bb z^\mu + \cc}{M^\#}\right),\
  \p_\mu\left(\frac{G z^\mu + \cc}{M^\#}\right),\
  \p_\mu\left(\frac{K\Bb z^\mu + \cc}{M^\#}\right);
\end{align}
with six derivatives and six fields
\begin{align}
  &\p_\mu\left(\frac{z^2\Hb z^{\mu\nu}\zb_\nu + \cc}{M^\#}\right),\
  \p_\mu\left(\frac{z^2\Hb\Bb z^\mu + \cc}{M^\#}\right),\
  \p_\mu\left(\frac{z^2\Jb z^\mu + \cc}{M^\#}\right),\non&
  \p_\mu\left(\frac{\zb^3 z^{\mu\nu}z_{\nu\rho}z^\rho + \cc}{M^\#}\right),\
  \p_\mu\left(\frac{\zb^2 K z^{\mu\nu}z_\nu + \cc}{M^\#}\right),\
  \p_\mu\left(\frac{\zb^2 H z^{\mu\nu}\zb_\nu + \cc}{M^\#}\right),\non&
  \p_\mu\left(\frac{\zb^2 I z^\mu + \cc}{M^\#}\right),\
  \p_\mu\left(\frac{\zb^3 E z^\mu + \cc}{M^\#}\right),\
  \p_\mu\left(\frac{\zb H\Hb z^\mu + \cc}{M^\#}\right),\
  \p_\mu\left(\frac{\zb^3 B^2 z^\mu + \cc}{M^\#}\right),\non&
  \p_\mu\left(\frac{z K\Hb z^\mu + \cc}{M^\#}\right),\
  \p_\mu\left(\frac{\zb^2 H\Bb z^\mu + \cc}{M^\#}\right),\
  \p_\mu\left(\frac{\zb K^2 z^\mu + \cc}{M^\#}\right),
\end{align}
are the complete lists of possibilities with no more than 3
derivatives acting on a single field.
With only two fields, the total derivatives become trivial and we will
not list them here.

\bibliographystyle{unsrturl}
\bibliography{refs}

\begin{thebibliography}{10}

\bibitem{Manohar:2018aog}
Aneesh~V. Manohar.
\newblock {Introduction to Effective Field Theories}.
\newblock 4 2018.
\newblock \href {http://arxiv.org/abs/1804.05863} {\path{arXiv:1804.05863}},
  \href {https://doi.org/10.1093/oso/9780198855743.003.0002}
  {\path{doi:10.1093/oso/9780198855743.003.0002}}.

\bibitem{Brivio:2017vri}
Ilaria Brivio and Michael Trott.
\newblock {The Standard Model as an Effective Field Theory}.
\newblock {\em Phys. Rept.}, 793:1--98, 2019.
\newblock \href {http://arxiv.org/abs/1706.08945} {\path{arXiv:1706.08945}},
  \href {https://doi.org/10.1016/j.physrep.2018.11.002}
  {\path{doi:10.1016/j.physrep.2018.11.002}}.

\bibitem{Scherer:2002tk}
Stefan Scherer.
\newblock {Introduction to chiral perturbation theory}.
\newblock {\em Adv. Nucl. Phys.}, 27:277, 2003.
\newblock \href {http://arxiv.org/abs/hep-ph/0210398}
  {\path{arXiv:hep-ph/0210398}}.

\bibitem{Epelbaum:2008ga}
Evgeny Epelbaum, Hans-Werner Hammer, and Ulf-G. Meissner.
\newblock {Modern Theory of Nuclear Forces}.
\newblock {\em Rev. Mod. Phys.}, 81:1773--1825, 2009.
\newblock \href {http://arxiv.org/abs/0811.1338} {\path{arXiv:0811.1338}},
  \href {https://doi.org/10.1103/RevModPhys.81.1773}
  {\path{doi:10.1103/RevModPhys.81.1773}}.

\bibitem{Machleidt:2011zz}
R.~Machleidt and D.~R. Entem.
\newblock {Chiral effective field theory and nuclear forces}.
\newblock {\em Phys. Rept.}, 503:1--75, 2011.
\newblock \href {http://arxiv.org/abs/1105.2919} {\path{arXiv:1105.2919}},
  \href {https://doi.org/10.1016/j.physrep.2011.02.001}
  {\path{doi:10.1016/j.physrep.2011.02.001}}.

\bibitem{Weinberg:1978kz}
Steven Weinberg.
\newblock {Phenomenological Lagrangians}.
\newblock {\em Physica A}, 96(1-2):327--340, 1979.
\newblock \href {https://doi.org/10.1016/0378-4371(79)90223-1}
  {\path{doi:10.1016/0378-4371(79)90223-1}}.

\bibitem{Ostrogradsky:1850fid}
M.~Ostrogradsky.
\newblock {M\'emoires sur les \'equations diff\'erentielles, relatives au
  probl\`eme des isop\'erim\`etres}.
\newblock {\em Mem. Acad. St. Petersbourg}, 6(4):385--517, 1850.

\bibitem{Woodard:2015zca}
Richard~P. Woodard.
\newblock {Ostrogradsky's theorem on Hamiltonian instability}.
\newblock {\em Scholarpedia}, 10(8):32243, 2015.
\newblock \href {http://arxiv.org/abs/1506.02210} {\path{arXiv:1506.02210}},
  \href {https://doi.org/10.4249/scholarpedia.32243}
  {\path{doi:10.4249/scholarpedia.32243}}.

\bibitem{Solomon:2017nlh}
Adam~R. Solomon and Mark Trodden.
\newblock {Higher-derivative operators and effective field theory for general
  scalar-tensor theories}.
\newblock {\em JCAP}, 02:031, 2018.
\newblock \href {http://arxiv.org/abs/1709.09695} {\path{arXiv:1709.09695}},
  \href {https://doi.org/10.1088/1475-7516/2018/02/031}
  {\path{doi:10.1088/1475-7516/2018/02/031}}.

\bibitem{Asorey:2020omv}
M.~Asorey, F.~Falceto, and L.~Rachwa\l{}.
\newblock {Asymptotic freedom and higher derivative gauge theories}.
\newblock {\em JHEP}, 05:075, 2021.
\newblock \href {http://arxiv.org/abs/2012.15693} {\path{arXiv:2012.15693}},
  \href {https://doi.org/10.1007/JHEP05(2021)075}
  {\path{doi:10.1007/JHEP05(2021)075}}.

\bibitem{Nicolis:2008in}
Alberto Nicolis, Riccardo Rattazzi, and Enrico Trincherini.
\newblock {The Galileon as a local modification of gravity}.
\newblock {\em Phys. Rev. D}, 79:064036, 2009.
\newblock \href {http://arxiv.org/abs/0811.2197} {\path{arXiv:0811.2197}},
  \href {https://doi.org/10.1103/PhysRevD.79.064036}
  {\path{doi:10.1103/PhysRevD.79.064036}}.

\bibitem{Horndeski:1974wa}
Gregory~Walter Horndeski.
\newblock {Second-order scalar-tensor field equations in a four-dimensional
  space}.
\newblock {\em Int. J. Theor. Phys.}, 10:363--384, 1974.
\newblock \href {https://doi.org/10.1007/BF01807638}
  {\path{doi:10.1007/BF01807638}}.

\bibitem{Skyrme:1961vq}
T.~H.~R. Skyrme.
\newblock {A Nonlinear field theory}.
\newblock {\em Proc. Roy. Soc. Lond. A}, 260:127--138, 1961.
\newblock \href {https://doi.org/10.1098/rspa.1961.0018}
  {\path{doi:10.1098/rspa.1961.0018}}.

\bibitem{Gudnason:2017opo}
Sven~Bjarke Gudnason and Muneto Nitta.
\newblock {A higher-order Skyrme model}.
\newblock {\em JHEP}, 09:028, 2017.
\newblock \href {http://arxiv.org/abs/1705.03438} {\path{arXiv:1705.03438}},
  \href {https://doi.org/10.1007/JHEP09(2017)028}
  {\path{doi:10.1007/JHEP09(2017)028}}.

\bibitem{Faddeev:1998eq}
L.~D. Faddeev and Antti~J. Niemi.
\newblock {Partially dual variables in SU(2) Yang-Mills theory}.
\newblock {\em Phys. Rev. Lett.}, 82:1624--1627, 1999.
\newblock \href {http://arxiv.org/abs/hep-th/9807069}
  {\path{arXiv:hep-th/9807069}}, \href
  {https://doi.org/10.1103/PhysRevLett.82.1624}
  {\path{doi:10.1103/PhysRevLett.82.1624}}.

\bibitem{Piette:1994jt}
B.~M. A.~G. Piette, W.~J. Zakrzewski, H.~J.~W. Mueller-Kirsten, and D.~H.
  Tchrakian.
\newblock {A Modified Mottola-Wipf model with sphaleron and instanton fields}.
\newblock {\em Phys. Lett. B}, 320:294--298, 1994.
\newblock \href {https://doi.org/10.1016/0370-2693(94)90659-9}
  {\path{doi:10.1016/0370-2693(94)90659-9}}.

\bibitem{Piette:1994ug}
B.~M. A.~G. Piette, B.~J. Schroers, and W.~J. Zakrzewski.
\newblock {Multi - solitons in a two-dimensional Skyrme model}.
\newblock {\em Z. Phys. C}, 65:165--174, 1995.
\newblock \href {http://arxiv.org/abs/hep-th/9406160}
  {\path{arXiv:hep-th/9406160}}, \href {https://doi.org/10.1007/BF01571317}
  {\path{doi:10.1007/BF01571317}}.

\bibitem{Anselmi:2017lia}
Damiano Anselmi and Marco Piva.
\newblock {Perturbative unitarity of Lee-Wick quantum field theory}.
\newblock {\em Phys. Rev. D}, 96(4):045009, 2017.
\newblock \href {http://arxiv.org/abs/1703.05563} {\path{arXiv:1703.05563}},
  \href {https://doi.org/10.1103/PhysRevD.96.045009}
  {\path{doi:10.1103/PhysRevD.96.045009}}.

\bibitem{Anselmi:2018tmf}
Damiano Anselmi and Marco Piva.
\newblock {Quantum Gravity, Fakeons And Microcausality}.
\newblock {\em JHEP}, 11:021, 2018.
\newblock \href {http://arxiv.org/abs/1806.03605} {\path{arXiv:1806.03605}},
  \href {https://doi.org/10.1007/JHEP11(2018)021}
  {\path{doi:10.1007/JHEP11(2018)021}}.

\bibitem{Anselmi:2019nie}
Damiano Anselmi and Antonio Marino.
\newblock {Fakeons and microcausality: light cones, gravitational waves and the
  Hubble constant}.
\newblock {\em Class. Quant. Grav.}, 37(9):095003, 2020.
\newblock \href {http://arxiv.org/abs/1909.12873} {\path{arXiv:1909.12873}},
  \href {https://doi.org/10.1088/1361-6382/ab78d2}
  {\path{doi:10.1088/1361-6382/ab78d2}}.

\bibitem{Antoniadis:2007xc}
I.~Antoniadis, E.~Dudas, and D.~M. Ghilencea.
\newblock {Supersymmetric Models with Higher Dimensional Operators}.
\newblock {\em JHEP}, 03:045, 2008.
\newblock \href {http://arxiv.org/abs/0708.0383} {\path{arXiv:0708.0383}},
  \href {https://doi.org/10.1088/1126-6708/2008/03/045}
  {\path{doi:10.1088/1126-6708/2008/03/045}}.

\bibitem{Dudas:2015vka}
E.~Dudas and D.~M. Ghilencea.
\newblock {Effective operators in SUSY, superfield constraints and searches for
  a UV completion}.
\newblock {\em JHEP}, 06:124, 2015.
\newblock \href {http://arxiv.org/abs/1503.08319} {\path{arXiv:1503.08319}},
  \href {https://doi.org/10.1007/JHEP06(2015)124}
  {\path{doi:10.1007/JHEP06(2015)124}}.

\bibitem{Fujimori:2016udq}
Toshiaki Fujimori, Muneto Nitta, and Yusuke Yamada.
\newblock {Ghostbusters in higher derivative supersymmetric theories: who is
  afraid of propagating auxiliary fields?}
\newblock {\em JHEP}, 09:106, 2016.
\newblock \href {http://arxiv.org/abs/1608.01843} {\path{arXiv:1608.01843}},
  \href {https://doi.org/10.1007/JHEP09(2016)106}
  {\path{doi:10.1007/JHEP09(2016)106}}.

\bibitem{Fujimori:2017rcc}
Toshiaki Fujimori, Muneto Nitta, Keisuke Ohashi, and Yusuke Yamada.
\newblock {Ghostbusters in $f(R)$ supergravity}.
\newblock {\em JHEP}, 05:102, 2018.
\newblock \href {http://arxiv.org/abs/1712.05017} {\path{arXiv:1712.05017}},
  \href {https://doi.org/10.1007/JHEP05(2018)102}
  {\path{doi:10.1007/JHEP05(2018)102}}.

\bibitem{Donoghue:2021eto}
John~F. Donoghue and Gabriel Menezes.
\newblock {Ostrogradsky instability can be overcome by quantum physics}.
\newblock {\em Phys. Rev. D}, 104(4):045010, 2021.
\newblock \href {http://arxiv.org/abs/2105.00898} {\path{arXiv:2105.00898}},
  \href {https://doi.org/10.1103/PhysRevD.104.045010}
  {\path{doi:10.1103/PhysRevD.104.045010}}.

\bibitem{Arici:2017whq}
Francesca Arici, Daniel Becker, Chris Ripken, Frank Saueressig, and Walter~D.
  van Suijlekom.
\newblock {Reflection positivity in higher derivative scalar theories}.
\newblock {\em J. Math. Phys.}, 59(8):082302, 2018.
\newblock \href {http://arxiv.org/abs/1712.04308} {\path{arXiv:1712.04308}},
  \href {https://doi.org/10.1063/1.5027231} {\path{doi:10.1063/1.5027231}}.

\bibitem{Gomis:1995jp}
Joaquim Gomis and Steven Weinberg.
\newblock {Are nonrenormalizable gauge theories renormalizable?}
\newblock {\em Nucl. Phys. B}, 469:473--487, 1996.
\newblock \href {http://arxiv.org/abs/hep-th/9510087}
  {\path{arXiv:hep-th/9510087}}, \href
  {https://doi.org/10.1016/0550-3213(96)00132-0}
  {\path{doi:10.1016/0550-3213(96)00132-0}}.

\bibitem{Anselmi:2013sx}
Damiano Anselmi.
\newblock {Renormalization of gauge theories without cohomology}.
\newblock {\em Eur. Phys. J. C}, 73:2508, 2013.
\newblock \href {http://arxiv.org/abs/1301.7577} {\path{arXiv:1301.7577}},
  \href {https://doi.org/10.1140/epjc/s10052-013-2508-5}
  {\path{doi:10.1140/epjc/s10052-013-2508-5}}.

\bibitem{Quadri:2021syf}
Andrea Quadri.
\newblock {Background field method and generalized field redefinitions in
  effective field theories}.
\newblock {\em Eur. Phys. J. Plus}, 136(6):695, 2021.
\newblock \href {http://arxiv.org/abs/2102.10656} {\path{arXiv:2102.10656}},
  \href {https://doi.org/10.1140/epjp/s13360-021-01665-9}
  {\path{doi:10.1140/epjp/s13360-021-01665-9}}.

\bibitem{Derrick:1964ww}
G.~H. Derrick.
\newblock {Comments on nonlinear wave equations as models for elementary
  particles}.
\newblock {\em J. Math. Phys.}, 5:1252--1254, 1964.
\newblock \href {https://doi.org/10.1063/1.1704233}
  {\path{doi:10.1063/1.1704233}}.

\bibitem{Gasser:1983yg}
J.~Gasser and H.~Leutwyler.
\newblock {Chiral Perturbation Theory to One Loop}.
\newblock {\em Annals Phys.}, 158:142, 1984.
\newblock \href {https://doi.org/10.1016/0003-4916(84)90242-2}
  {\path{doi:10.1016/0003-4916(84)90242-2}}.

\bibitem{Bijnens:1999sh}
Johan Bijnens, Gilberto Colangelo, and Gerhard Ecker.
\newblock {The Mesonic chiral Lagrangian of order p**6}.
\newblock {\em JHEP}, 02:020, 1999.
\newblock \href {http://arxiv.org/abs/hep-ph/9902437}
  {\path{arXiv:hep-ph/9902437}}, \href
  {https://doi.org/10.1088/1126-6708/1999/02/020}
  {\path{doi:10.1088/1126-6708/1999/02/020}}.

\bibitem{Bijnens:2018lez}
Johan Bijnens, Nils Hermansson-Truedsson, and Si~Wang.
\newblock {The order p$^{8}$ mesonic chiral Lagrangian}.
\newblock {\em JHEP}, 01:102, 2019.
\newblock \href {http://arxiv.org/abs/1810.06834} {\path{arXiv:1810.06834}},
  \href {https://doi.org/10.1007/JHEP01(2019)102}
  {\path{doi:10.1007/JHEP01(2019)102}}.

\bibitem{Graf:2020yxt}
Lukas Graf, Brian Henning, Xiaochuan Lu, Tom Melia, and Hitoshi Murayama.
\newblock {2, 12, 117, 1959, 45171, 1170086, \textellipsis{}: a Hilbert series
  for the QCD chiral Lagrangian}.
\newblock {\em JHEP}, 01:142, 2021.
\newblock \href {http://arxiv.org/abs/2009.01239} {\path{arXiv:2009.01239}},
  \href {https://doi.org/10.1007/JHEP01(2021)142}
  {\path{doi:10.1007/JHEP01(2021)142}}.

\bibitem{Motohashi:2014opa}
Hayato Motohashi and Teruaki Suyama.
\newblock {Third order equations of motion and the Ostrogradsky instability}.
\newblock {\em Phys. Rev. D}, 91(8):085009, 2015.
\newblock \href {http://arxiv.org/abs/1411.3721} {\path{arXiv:1411.3721}},
  \href {https://doi.org/10.1103/PhysRevD.91.085009}
  {\path{doi:10.1103/PhysRevD.91.085009}}.

\bibitem{Motohashi:2016ftl}
Hayato Motohashi, Karim Noui, Teruaki Suyama, Masahide Yamaguchi, and David
  Langlois.
\newblock {Healthy degenerate theories with higher derivatives}.
\newblock {\em JCAP}, 07:033, 2016.
\newblock \href {http://arxiv.org/abs/1603.09355} {\path{arXiv:1603.09355}},
  \href {https://doi.org/10.1088/1475-7516/2016/07/033}
  {\path{doi:10.1088/1475-7516/2016/07/033}}.

\bibitem{Crisostomi:2017aim}
Marco Crisostomi, Remko Klein, and Diederik Roest.
\newblock {Higher Derivative Field Theories: Degeneracy Conditions and
  Classes}.
\newblock {\em JHEP}, 06:124, 2017.
\newblock \href {http://arxiv.org/abs/1703.01623} {\path{arXiv:1703.01623}},
  \href {https://doi.org/10.1007/JHEP06(2017)124}
  {\path{doi:10.1007/JHEP06(2017)124}}.

\bibitem{Motohashi:2017eya}
Hayato Motohashi, Teruaki Suyama, and Masahide Yamaguchi.
\newblock {Ghost-free theory with third-order time derivatives}.
\newblock {\em J. Phys. Soc. Jap.}, 87(6):063401, 2018.
\newblock \href {http://arxiv.org/abs/1711.08125} {\path{arXiv:1711.08125}},
  \href {https://doi.org/10.7566/JPSJ.87.063401}
  {\path{doi:10.7566/JPSJ.87.063401}}.

\bibitem{Motohashi:2018pxg}
Hayato Motohashi, Teruaki Suyama, and Masahide Yamaguchi.
\newblock {Ghost-free theories with arbitrary higher-order time derivatives}.
\newblock {\em JHEP}, 06:133, 2018.
\newblock \href {http://arxiv.org/abs/1804.07990} {\path{arXiv:1804.07990}},
  \href {https://doi.org/10.1007/JHEP06(2018)133}
  {\path{doi:10.1007/JHEP06(2018)133}}.

\bibitem{Ganz:2020skf}
Alexander Ganz and Karim Noui.
\newblock {Reconsidering the Ostrogradsky theorem: Higher-derivatives
  Lagrangians, Ghosts and Degeneracy}.
\newblock {\em Class. Quant. Grav.}, 38(7):075005, 2021.
\newblock \href {http://arxiv.org/abs/2007.01063} {\path{arXiv:2007.01063}},
  \href {https://doi.org/10.1088/1361-6382/abe31d}
  {\path{doi:10.1088/1361-6382/abe31d}}.

\bibitem{Grosse-Knetter:1993tae}
Carsten Grosse-Knetter.
\newblock {Effective Lagrangians with higher derivatives and equations of
  motion}.
\newblock {\em Phys. Rev. D}, 49:6709--6719, 1994.
\newblock \href {http://arxiv.org/abs/hep-ph/9306321}
  {\path{arXiv:hep-ph/9306321}}, \href
  {https://doi.org/10.1103/PhysRevD.49.6709}
  {\path{doi:10.1103/PhysRevD.49.6709}}.

\end{thebibliography}

\end{document}